\documentclass[numberedappendix]{emulateapj}

\def\MyApJNew#1{#1}
\def\MyMNRASNew#1{}
\MyMNRASNew{\RequirePackage{rotating}}

\usepackage{graphicx}
\usepackage{dcolumn}
\usepackage{bm}
\usepackage{epsf}
\usepackage{verbatim}
\usepackage{amsmath}
\usepackage{amssymb}
\usepackage{amsfonts}
\usepackage{ifthen}
\usepackage{epsfig}
\usepackage{verbatim}
\usepackage{dcolumn}
\usepackage{epstopdf}
\usepackage{subfigure}
\usepackage{threeparttable}
\usepackage{booktabs}
\usepackage{calc}

\usepackage{multirow}
\usepackage{comment}
\MyMNRASNew{\usepackage{lscape}}

\usepackage[figuresright]{rotating}

\def\myfig#1{./Figures/#1}
\def\mybib#1{./#1}

\def\DrawFig#1{#1}

\newcommand{\Zjump}{q_{\text{\fontsize{5}{6}\selectfont{\mbox{$Z$}}}}}
\newcommand{\Kjump}{q_{\text{\fontsize{5}{6}\selectfont{\mbox{$K$}}}}}

\newcommand{\lc}{{\Delta r}}

\newcommand{\rT}{r_{\text{\fontsize{5}{6}\selectfont T}}}

\newcommand{\Myfi}{{f_i}}
\newcommand{\Myfo}{{f_o}}
\newcommand{\MyfK}{{f_K}}
\newcommand{\MyfZ}{{f_Z}}
\newcommand{\MyfKi}{{f_{Ki}}}
\newcommand{\MyfKo}{{f_{Ko}}}
\newcommand{\MyfZi}{{f_{Zi}}}
\newcommand{\MyfZo}{{f_{Zo}}}

\newcommand{\barMyfK}{{\bar{f}_K}}
\newcommand{\barMyfZ}{{\bar{f}_Z}}
\newcommand{\barMyfKi}{{\bar{f}_{Ki}}}
\newcommand{\barMyfKo}{{\bar{f}_{Ko}}}

\newcommand{\qrZ}{\MyfZ}
\newcommand{\qrK}{\MyfK}
\newcommand{\barqrZ}{\barMyfZ}
\newcommand{\barqrK}{\barMyfK}

\newcommand{\rKircf}{\MyfKi}
\newcommand{\rKorcf}{\MyfKo}
\newcommand{\rZircf}{\MyfZi}
\newcommand{\rZorcf}{\MyfZo}

\newcommand{\barrKircf}{\barMyfKi}
\newcommand{\barrKorcf}{\barMyfKo}

\newcommand{\Sx}{S_x}

\newcommand{\rcf}{r_{\tiny \text{cf}}}
\newcommand{\rR}{r_{\tiny \text{0}}}
\newcommand{\Rcf}{r_{\tiny \text{cf}}}
\newcommand{\riZ}{r_{\tiny \text{Z,i}}}
\newcommand{\roZ}{r_{\tiny \text{Z,o}}}
\newcommand{\riK}{r_{\tiny \text{K,i}}}
\newcommand{\roK}{r_{\tiny \text{K,o}}}

\newcommand{\tdriZ}{\tilde{r}_{\tiny \text{Z,i}}}
\newcommand{\tdroZ}{\tilde{r}_{\tiny \text{Z,o}}}
\newcommand{\tdriK}{\tilde{r}_{\tiny \text{K,i}}}
\newcommand{\tdroK}{\tilde{r}_{\tiny \text{K,o}}}

\newcommand{\TtwoD}{T_{\text{\fontsize{5}{6}\selectfont proj}}}
\newcommand{\ZtwoD}{Z_{\text{\fontsize{5}{6}\selectfont proj}}}

\newcommand{\bT}{b_{\text{\fontsize{5}{6}\selectfont T}}}
\newcommand{\cT}{c_{\text{\fontsize{5}{6}\selectfont T}}}

\newcommand{\Macht}{\mathcal{M}}

\newcommand{\alphaT}{\alpha_{\tiny \text{T}}}

\newcommand{\alphaZ}{\alpha_{\tiny \text{Z}}}

\newcommand{\corl}{\mathcal{R}}

\newcommand{\DrOZZ}{\Delta_{100}}

\newcommand{\tdyn}{t_{\tiny\text{dyn}}}
\newcommand{\tcool}{t_{\tiny\text{cool}}}
\newcommand{\theat}{t_{\tiny\text{heat}}}
\newcommand{\tcond}{t_{\tiny\text{cond}}}
\newcommand{\tdiff}{t_{\tiny\text{diff}}}

\newcommand{\tmix}{t_{\tiny\text{mix}}}
\newcommand{\tadv}{t_{\tiny\text{adv}}}
\newcommand{\tadvF}{t_{\tiny\text{adv,f}}}

\newcommand{\rad}{\mbox{ rad}}
\newcommand{\BmuG}{B_{\mu\text{G}}}

\MyApJNew{\bibliographystyle{apj}}
\MyMNRASNew{\bibliographystyle{mnras}}

\defcitealias{ReissKeshet2014}{RK14}

\defcitealias{KeshetEtAl2010}{K10}

\defcitealias{Keshet2012}{K12}

\defcitealias{NaorEtal2020}{N20}
\newcommand{\NaorEAIP}{{\citetalias{NaorEtal2020}}}

\newcommand{\ie}{\emph{i.e.} }
\newcommand{\eg}{\emph{e.g.,} }

\newcommand{\be}{\begin{equation}}
\newcommand{\ee}{\end{equation}}
\newcommand{\bea}{\begin{equation*}}
\newcommand{\eea}{\end{equation*}}
\newcommand{\beqr}{\begin{eqnarray} \nonumber}
\newcommand{\eeqr}{\end{eqnarray}}
\newcommand{\beqrb}{\begin{eqnarray}}
\newcommand{\eeqrb}{\nonumber \end{eqnarray}}
\newcommand{\fin}{\mbox{ .}}
\newcommand{\coma}{\mbox{ ,}}

\newcommand{\cm}{\mbox{ cm}}

\newcommand{\se}{\mbox{ s}}

\newcommand{\Gyr}{\mbox{ Gyr}}
\newcommand{\erg}{\mbox{ erg}}

\newcommand{\kpc}{\mbox{ kpc}}

\newcommand{\keV}{\mbox{ keV}}

\newcommand{\muG}{\mbox{ $\mu$G}}

\newcommand{\const}{\mbox{const.}}

\newcommand{\vect}[1]{\mathbf{#1}}

\usepackage{xcolor}
\definecolor{darkgreen}{rgb}{0.0,0.5,0.0}
\MyApJNew{\usepackage[colorlinks,citecolor=darkgreen]{hyperref} 
}

\MyApJNew{}

\MyApJNew{
\begin{document}}

\MyApJNew{
\title{Advection by large-scale spiral flows in galaxy clusters
}
\shorttitle{Spiral advection in GCs}
}

\MyApJNew{
    \shortauthors{Naor et al.}
	\author{Yossi Naor$^{\dagger}$}
	\author{Uri Keshet$^{\ddagger}$}
	\affil{Physics Department, Ben-Gurion University of the Negev, POB 653, Be'er-Sheva 84105, Israel}
	\thanks{$^{\dagger}$Electronic address: naoryos@post.bgu.ac.il}	
	\thanks{$^{\ddagger}$Electronic address: ukeshet@bgu.ac.il}
	}
	
	\MyMNRASNew{
    \title[Spiral advection in GCs]{Advection by large scale spiral flows in galaxy clusters}
	\author[Naor et al.]{Yossi Naor, \&
	Uri Keshet\thanks{E-mail: ukeshet@bgu.ac.il}
	\\
	Physics Department, Ben-Gurion University of the Negev, POB 653, Be'er-Sheva 84105, Israel\\
	}
	}
	
\date{\today}

\MyMNRASNew{
\pubyear{2020}
\begin{document}
\label{firstpage}
\pagerange{\pageref{firstpage}--\pageref{lastpage}}
\maketitle
}	
	
\begin{abstract}
The intracluster medium of a galaxy cluster often shows an extended quasi-spiral structure, accentuated by tangential discontinuities known as cold fronts (CFs).
These discontinuities are thought to isolate between low-entropy, high-metallicity gas inside (\emph{i.e.}, below) the CF that was lifted from the center of the cluster by some radial factor $f_i$, and high-entropy, low-metallicity gas outside the CF that was pushed inward by a factor $f_o$.
We find broad support for such a picture, by comparing the entropy and metallicity discontinuities with the respective azimuthal averages, using newly deprojected thermal profiles in clusters A2029, A2142, A2204 and Centaurus, supplemented by deprojected CFs from the literature.
In particular, the mean advection factors $f_K$ and $f_Z$, inferred respectively from entropy and metallicity, strongly correlate ($\mathcal{R}=0.7^{+0.2}_{-0.3}$) with each other, consistent with large-scale advection.
However, unlike sloshing simulations, in which the inside/outside phases are an inflow/outflow settling back to equilibrium after a violent perturbation, our results are more consistent with an outflow/inflow, with the fast, Mach $\mathcal{M}_i\sim0.8$ gas inside the CF being a rapidly heated or mixed outflow, probably originating from the cD galaxy, and gas outside the CF being a $\mathcal{M}_o\sim0.03$, slowly-cooling inflow.
In particular, entropy indicates an outside advection factor $f_{Ko}\simeq 1.33\pm0.04$
that is approximately constant in all CFs, gauging the distance traversed by inflowing gas within a cooling time.
In contrast, $1.1\lesssim f_{Ki}\lesssim 2.5$ and $1\lesssim f_Z\lesssim 17$ vary considerably among clusters, and strongly correlate ($3.1\sigma\mbox{--}4.2\sigma$) with the virial mass, $f_{Ki}\propto M_{200}^{0.14\pm0.07}$ and $f_Z\propto M_{200}^{1.4\pm0.4}$, suggesting that each cluster sustains a quasi-steady spiral flow.
\end{abstract}

\date{Accepted ---. Received ---; in original ---}
	
\MyApJNew{\label{firstpage}}

\MyMNRASNew{
}
	
\MyApJNew{	
}

\MyApJNew{	
\maketitle
}

\section{Introduction}
\label{sec:Introduction_Mixing}

During the past two decades, high resolution X-ray imaging of the intracluster medium (ICM) of galaxy clusters (GCs) has revealed an abundance of thermal discontinuities known as cold fronts \citep[CFs; for a review, see][]{MarkevitchVikhlinin2007}.
These CFs have proved to be useful diagnostics of the ICM.
As one crosses such a CF outward (\ie with increasing distance from the center of the GC), the plasma density sharply decreases (henceforth: drops) and its temperature sharply increases (henceforth: jumps), over a length scale shorter than the Coulomb mean free path \citep[\eg][]{EttoriFabian2000,VikhlininEtAl2001,WernerEtAl2016B}.
CFs are broadly classified into spiral-type and merger-type categories \citep[\eg][]{MarkevitchVikhlinin2007}, the former presenting a quasi-spiral structure, and the latter showing a bullet-like or other unrelaxed morphology associated with a recent merger event.
Other, putative types of CFs \citep[\eg][]{BirnboimEtAl10} were not yet detected.
We henceforth focus exclusively on spiral CFs.

Spiral-type CFs are tangential discontinuities, widely believed to arise from large-scale, sloshing motions of the ICM following some violent trigger event.
Such CFs show hydrostatic mass jumps that reveal nearly sonic tangential flows below the CFs \citep{KeshetEtAl2010}, with a typical Mach number $\Macht\sim0.8$ \citep[][henceforth \NaorEAIP]{NaorEtal2020}, and thermal pressure jumps that indicate magnetization levels enhanced by $\sim 10\%$ below the CFs, probably due to shear \citep[][ and \NaorEAIP]{ReissKeshet2014}.
Sloshing may be driven by mergers, possibly involving only a dark matter subhalo \citep[\eg][]{TittleyHenriksen2005, AscasibarMarkevitch2006, ZuHoneEtAl2010, RoedigerEtAl2011}, or by weak shocks or acoustic waves that are propagating towards the GC center \citep{ChurazovEtAl2003,FujitaEtAl2004}. They may be a feature of a long-lived spiral flow, shaping the cluster core, not necessarily associated with a single violent event \citep{Keshet2012}.

Simulations of ICM sloshing are typically triggered by a merger event \citep[\eg][]{AscasibarMarkevitch2006, ZuHoneEtAl2010, RoedigerEtAl2011, ZuhoneEtAl2013, UedaEtAl2019} and are usually adiabatic \citep[see however][for non-adiabatic sloshing simulations]{ZuHoneEtAl2010}.
The resulting spiral structure consists of cold gas violently lifted from the central regions of the cluster and gradually falling back to the center, and hot gas that was initially trapped near the center and gradually rises back to its origin \citep{Keshet2012}.
The simulated CF is thus found to isolate between two phases: an inner phase lifted from the center by some radial factor $1<\Myfi\lesssim 2$, and hot gas lowered from radii larger by a comparable factor, $\Myfo\simeq \Myfi$ \citep{AscasibarMarkevitch2006}; subscript $i$ ($o$) denotes regions inside (outside) the CF, namely below (above) it.
However, in the presence of radiative cooling, the hot phase is more likely to be an inflow \citep{Keshet2012}.
Incorporating radiative cooling in a merger-triggered sloshing simulation leads to a cooling flow, only mildly slowed down by sloshing \citep[][]{ZuHoneEtAl2010}.

The aforementioned drops in density and jumps in temperature, pressure, and hydrostatic mass, reflect the underlying ICM dynamics, evolving over the relatively short, dynamical timescale.
For a discussion of such dynamical characteristics of the spiral structure, its manifestation in CFs, and its underlying flow, see \NaorEAIP.
In this paper, we focus on diagnostics of the CFs and spiral structure that evolve much more slowly, over timescales longer than the dynamical time, thus revealing the long-term evolution of the ICM.
By studying such diagnostics, one can identify the role of additional physical processes, in particular non-adiabatic heating and cooling, and constrain the mechanism giving rise to the spiral structure and the underlying dynamics.

One such slowly-evolving diagnostic is the entropy of a (Lagrangian) gas parcel, stable over a relatively long timescale dictated by radiative cooling, non-adiabatic heating, or heat conduction.
The drop in electron density $n$ and the jump in the electron temperature $T$, as one crosses outside a CF, imply a substantial coincident  jump in entropy, which we quantify by the adiabat $K= n^{-2/3}k_BT$ (hereafter referred to simply as entropy), where $k_B$ is the Boltzmann constant.
As we show below, tracing the entropy jumps across CFs constrains the ICM evolution over the recent $\lesssim \mbox{Gyr}$ timescale.

Another slowly-evolving diagnostic is the metallicity $Z$ of a gas parcel, which typically varies in the ICM over even longer timescales, associated with the expansion of galactic winds and with turbulent diffusion.
In spite of the challenges involved in deprojecting the metallicity profile, sharp drops in metallicity were recently established across CFs \citep[][\NaorEAIP]{SandersEtAl2005,SandersEtAl2016}.
As we show below, metallicity gradients over the relevant, spiral lengthscales, evolve over long timescales, of several Gyr or more.
Therefore, tracing the metallicity drops across CFs constrains the long-term evolution of the ICM.
If some violent, \eg merger, event triggered the formation of the spiral in the past few Gyr, a metallicity map can directly trace the original gas distribution.

Combining these entropy and metallicity tracers provides a useful tool for testing and constraining models for the origin and evolution of the spiral structure and its underlying flow.
In particular, consider a cluster with some well-defined declining radial profile $Z(r)$ and rising profile $K(r)$, that provide a good overall description of the ICM, at least at an early stage before the onset of the spiral flow.
If the Lagrangian metallicity can be approximated as time-independent, then measuring the metallicity $Z_i$ just below a CF yields the original radius $r_{i}$ of the fluid element, through $Z(r_{i})\simeq Z_i$. The original radius of the fluid element just above the CF can be similarly inferred from $Z(r_o)\simeq Z_o$. As long as the Lagrangian entropy remains approximately time-independent, $r_i$ and $r_o$ can be analogously inferred from $K_i$, $K_o$, and $K(r)$.
We henceforth denote values based on metallicity (on entropy) by subscript $Z$ ($K$).

In a young spiral structure, of an age $\lesssim \Gyr$ shorter than the entropy and diffusive timescales, metallicity and entropy should independently agree on the same origin of the phase below the CF, and similarly for the phase above the CF.
Namely, $\riK\simeq \riZ< \rcf$ should point at the same inner radius, below the CF radius $\rcf$, from which gas has been lifted to $\rcf$.
Similarly, $\roK\simeq \roZ>\rcf$ should similarly reflect the same outer radius, from which gas was lowered to $\rcf$.
In an older, $\gtrsim\Gyr$ spiral, $K$ will become increasingly modified by entropy-changing processes, but $Z$ will still be approximately preserved along the flow.
Consequently, differences will develop between the two inner radii, $\riK$ and $\riZ$, and between the two outer radii, $\roK$ and $\roZ$.
Nevertheless, some correlations within each pair should persist in this picture, in which the CF entropy jump and metallicity drop arise from the same dynamical mechanism.
Note that the evolution in $\riK$ is particularly sensitive, even over $\lesssim \Gyr$
timescales, to the level at which the local thermal instability is quenched.

We study the entropy and metallicity distributions in GCs with spiral structure, in order to constrain the underlying flow and its history.
Radial entropy and metallicity profiles are extracted from \emph{Chandra} data, by deprojecting the thermal properties along the line of sight.
For this purpose, we select eight spiral-type CFs in four GCs: one CF in Abell 2029 (A2029, henceforth), one in A2142, four in A2204, and two in the Centaurus cluster.
By deprojecting the data in angular sectors harbouring these CFs, we deduce the $K$ and $Z$ values just above and just below each CF.
These newly deprojected CFs are supplemented by a sample of eight high-quality deprojected CFs from the literature.

GCs show rather universal radial profiles of entropy \citep[$K\propto r$; \eg][]{TozziNorman2001,VoitEtAl2005,PanagouliaEtAl2014,ReissKeshet2015} and metallicity \citep[$Z\propto r^{-0.3}$; \eg][]{SandersonEtAl2009}, when properly rescaled. This universality suggests that the aforementioned, unperturbed profiles $K(r)$ and $Z(r)$ can be approximately determined. As a proxy of these profiles, we measure the (full-angle) azimuthal averages of $K$ and $Z$ in each of the analyzed clusters.
Such an approximation is good, provided that the flow and its trigger mechanism do not introduce substantial non-adiabatic or chemical changes in the overall, azimuthally averaged, properties of the ICM.

The paper is \MyMNRASNew{organized}\MyApJNew{organized} as follows.
In \S\ref{sec:TimeScales_Mixing}, we evaluate the timescales of key physical processes operating in the ICM.
In \S\ref{sec:Data_Red_Spect_An_Mixing}, we describe the data reduction, spectral analysis, and models used to deproject the thermal profiles.
The analysis of advection toward spiral CFs is presented in \S\ref{sec:advection_results_Mixing}. Our results are summarized and discussed in \S\ref{sec:Discussion_Mixing}.
We adopt a concordance $\Lambda$CDM model with a Hubble parameter $H_0=70\,\mbox{km}\,\mbox{s}^{-1}\,\mbox{Mpc}^{-1}$
and a matter fraction $\Omega_m=0.3$.
An angular separation of $1''$ is then equivalent to a proper-distance separation of $1.45\kpc$ in A2029, $2.65\kpc$ in A2204, $1.68\kpc$ in A2142, and $0.22\kpc$ in Centaurus.
A $76\%$ hydrogen mass fraction is assumed. Errors are quoted at the single-parameter $1\sigma$ confidence level, unless otherwise stated.

\section{Processes affecting the spiral structure
}\label{sec:TimeScales_Mixing}

Before proceeding to analyze the advection implied by the entropy and metallicity profiles, we evaluate the relevance of $K$ and $Z$ as tracers of the long-term evolution of the ICM.
For this purpose, we examine the typical timescales of physical processes in the ICM that can substantially modify $K$ and $Z$, and compare them to the short, dynamical timescale that governs pressure balance in the flow.
We evaluate these timescales as a function of radius $r$, measured with respect to the center of the GC, with an emphasis on the analyzed CF regions, as illustrated in Figure \ref{fig:timescales_Mixing}.

Denote the characteristic radial lengthscale for thermal variations in the spiral by $\lc$, which is of order $r$.
The dynamical timescale can be defined as the sound crossing time,
\begin{equation}\label{eq:t_dyn_Mixing}
\tdyn \equiv\frac{\lc}{c_s}\simeq0.1\,\DrOZZ T_3^{-\frac{1}{2}}\mbox{ Gyr}\coma
\end{equation}
where $\lc\equiv 100\DrOZZ\kpc$, $c_{s}=(\gamma k_B T/\mu)^{1/2}$ is the speed of sound, $\gamma=5/3$ is the adiabatic index, $\mu\simeq0.6m_p$ is the average particle mass, $m_p$ is the proton mass, and $k_B T\equiv3T_3\keV$.

This timescale is similar to the advection time along the spiral, given the aforementioned evidence for $\Macht\simeq 0.8$ flows just below the CF.
These flows are oriented along the spiral pattern, which typically subtends an angle $\alpha\simeq 0.2\rad$ with respect to the azimuth \citep{Keshet2012}, so the advection time by the fast flow is
\begin{equation}\label{eq:t_advf_Mixing}
\tadvF\sim(\Macht\sin\alpha)^{-1}\tdyn\simeq 6\tdyn\fin
\end{equation}

\MyApJNew{\begin{figure}[h]}
\MyMNRASNew{\begin{figure}}
	\centering
	\DrawFig{\includegraphics[width=8.6cm]{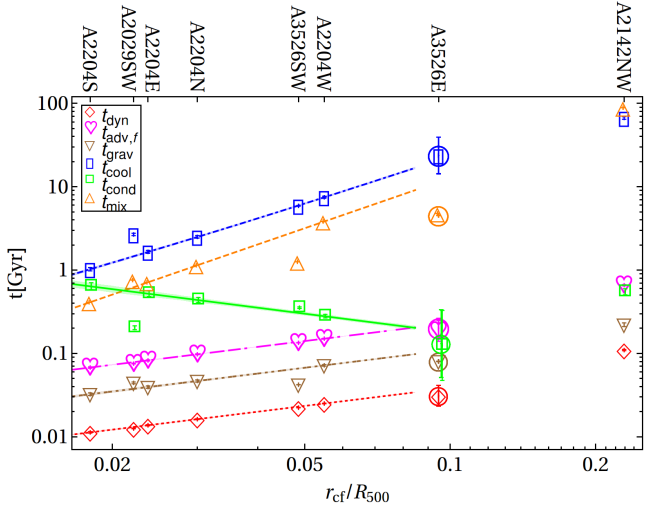}}
	\caption{
Timescales of key processes in GC spirals, derived based on deprojected, azimuthally-averaged, radial profiles, in A2204 (curves with narrow, shaded $1\sigma$ confidence levels) and near all CFs in our sample (symbols with error bars, labeled in upper axis), plotted against the normalized radius $\Rcf/R_{500}$.
Timescales include the dynamical time (Eq.~\ref{eq:t_dyn_Mixing}; red diamonds and dotted curve), advection time of the fast flow (Eq.~\ref{eq:t_advf_Mixing}; magenta hearts and dashed-dashed-dot curve), gravitational time (Eq.~\ref{eq:t_grav_Mixing}; brown down-pointing triangles and dot-dot-dashed curve), cooling time (Eq.~\ref{eq:t_cooling_Mixing}; blue rectangle and dot-dashed curve), heat conduction time (Eq.~\ref{eq:t_heat_cond_Mixing}, green squares and solid curve), and effective mixing time (Eq.~\ref{eq:t_turb_Mixing}; orange up-pointing triangles and dashed curve), all evaluated for a radial scale $\lc=\Rcf/2$. Circled symbols indicate extrapolated values from the global profile.
To show all symbols, some results are slightly shifted in abscissa (here and in subsequent figures).
}
	\label{fig:timescales_Mixing}
\end{figure}

Another timescale, which may regulate the ICM dynamics in some sloshing scenarios, is the gravitational time, which we define as
\begin{equation}\label{eq:t_grav_Mixing}
t_{\tiny\text{grav}}\equiv\sqrt{\frac{r^3}{G M(r)}}\simeq0.5 r_{100}^{\frac{3}{2}}M_{12}^{-\frac{1}{2}}\mbox{ Gyr}\coma
\end{equation}
where $r\equiv 100r_{100}\kpc$.
We assume that the hydrostatic mass,
\begin{equation}
M_h(r)\equiv10^{12} M_{12}\,M_\sun\equiv -\frac{r^2}{\mu n G}\partial_r(n\,k_BT) \coma
\end{equation}
where $G$ is the gravitational constant, provides an adequate proxy for the gravitating mass $M(r)$ for any $r$.
This assumption is inaccurate if $M_h$ is inferred from $n(r)$ and $T(r)$ in a narrow angular sector that harbors a nearly sonic flow, where substantial discrepancies between $M$ and $M_h$ are indeed identified, but it provides a reasonable approximation if the sector is sufficiently wide.

The flow underlying GC spiral structures is found to be sub-sonic, as expected from an apparently long-lived, stable configuration.
Entropy is mainly regulated, therefore, by radiative cooling and non-adiabatic heating.
Various heating mechanisms were invoked to explain the absence of cold gas and of excessive star formation in the core, in spite of the fast cooling \citep[\eg][]{Fabian1994, HudsonEtAl2010, ZhangEtAl2016,McDonaldEtAl2018}.
Suggestions include thermal conduction \citep[\eg][]{NarayanMedvedev2001, ZakamskaNarayan2003} and self regulating AGN activity \citep[\eg][]{GaspariEtAl2013, MartizziEtAl2019}, as well as sound waves, shocks, cosmic-rays, viscous heating, turbulent heating due to mergers, and galaxy motion \citep[for reviews, see][]{McNamaraNulsen2012, FabianEtAl2017, RuszkowskiEtAL2017}.
However, no mechanism was shown to quench both global and local thermal instabilities, yet maintain the delicate spiral structure and universal entropy profile, both of which suggest a dynamical heating mechanism \citep{ReissKeshet2015}.

As the core appears to show a balance between heating and radiative cooling, and the dominant heating process is yet to be identified, we assume that the heating and cooling times are comparable $\theat\sim\tcool$, and focus on the cooling time,
\begin{equation}\label{eq:t_cooling_Mixing}
t_{\tiny\text{cool}}\equiv\frac{u_{th}}{j}\simeq \frac{\zeta nk_BT}{(\gamma-1)j_x}\simeq40\,T_3 n_3^{-1}\Lambda_{23}^{-1}\mbox{Gyr}\coma
\end{equation}
where $u_{th}$ is the thermal energy density, $j$ and $j_x=n^2\Lambda$ are the bolometric and X-ray emissivities, $\Lambda\equiv10^{-23}\Lambda_{23}\erg \cm^{3}\se^{-1}$ is the cooling function, $\zeta\simeq1.9$ is the ratio between particle number density and electron number density, and $n\equiv 10^{-3}n_3\cm^{-3}$.
For a given GC redshift $z$, observed photon energy band $\{\epsilon_a,\epsilon_b\}$, and given gas parameters $T$ and $Z$, the spectral analysis package XSPEC\footnote{https://heasarc.gsfc.nasa.gov/xanadu/xspec} provides the cooling function $\Lambda$.
We model $\Lambda$ by interpolating tabular data from XSPEC, for the given GC redshift and the observed energy band, \ie $\epsilon_a=0.8\,\keV$ and $\epsilon_b=7.0\,\keV$.

In the spiral structure, hot and cold phases are brought to relative proximity, so it is useful to estimate the timescale for radial heat conduction between these components. Naively, invoking the Spitzer thermal conductivity \citep{Spitzer1962},
$\kappa_s\simeq2.9\times10^{14}\,T_3^{5/2}/\ln(\Lambda_S) \erg\se^{-1}\cm^{-1}\mbox{ K}^{-1}$,
one finds a very short heat conduction time,
\begin{equation} \label{eq:tcond_Mixing}
\tcond^{(s)} \simeq \frac{l_T^2}{2D_{\kappa}}\simeq \frac{nC_\text{v}\lc^2}{2\kappa_s}\simeq 0.07\frac{n_3\DrOZZ^2L_S}{T_3^{5/2}}\mbox{Gyr} \fin
\end{equation}
Here, $D_\kappa$ is the thermal diffusivity, $C_\text{v}=\gamma k_B/(\gamma-1)$ is the heat capacity at constant volume, and
\begin{equation}\label{eq:Lambda_spitzer_Mixing}
\ln(\Lambda_S)\simeq38.9+\ln\left(T_3^{3/2}n_3^{-1/2}\right)\equiv 38.9L_S
\end{equation}
is the Coulomb logarithm \citep[][]{Spitzer1962}.
The radial scale $l_T$, corresponding to order unity changes in temperature, is typically quite large in GCs, but is much smaller in a spiral structure. For a logarithmic spiral, which provides a reasonable approximation for observations \citep{Keshet2012}, $l_T\simeq r$.

However, strong magnetization, at the $\sim 10\%$ equipartition level, has been identified below CFs, likely associated with fields largely oriented along the discontinuity surface due to shear amplification \citep[see][and \NaorEAIP]{ZuHoneEtAl2011,ReissKeshet2014}.
Such magnetic field lines can suppress the radial heat conduction by many orders of magnitude \citep[\eg][]{RosenbluthKaufman1958}.
Assuming that the magnetic field nevertheless has a sufficiently tangled component, one can approximate the radial conduction by substituting $l_c$ in Eq.~(\ref{eq:tcond_Mixing}) with its random walk estimate, $l_c^2/l_{\tiny\text{B}}$, where $l_{\tiny\text{B}}\equiv 10 l_{\tiny\text{B},10}\kpc$ is the typical magnetic coherence length in the ICM \citep[estimated by, \eg][and references therein]{CarilliTaylor2002, SubramanianEtAl2006, BohringerEtAl2016}.
The radial heat-conduction time thus becomes \citep[\eg][]{RechesterRosenbluth1978,NarayanMedvedev2001}
\begin{equation}\label{eq:t_heat_cond_Mixing}
\tcond \simeq
6.9\frac{n_3\DrOZZ^4L_S}{T_3^{5/2}l^2_{\tiny\text{B},10}}\mbox{ Gyr}\fin
\end{equation}
Note the strong scale dependence; below we take $l_{\tiny\text{B}}\propto \Delta r$ and $\lc = f_\Delta r$, such that $\tcond\propto f_\Delta^4 r^2$.

Next, consider the processes governing changes in ICM metallicity over $\gtrsim 10\kpc$, spiral scales.
Assuming that the effects of galaxies are not correlated with the spiral structure, we focus on particle diffusion.
The diffusion time can be estimated as \citep[\eg][]{Spitzer1962}
\begin{equation}\label{eq:t_viscosity_Mixing}
\tdiff \sim\frac{\lc^2}{v\lambda}
>\frac{\lc^2}{v\,r_L}
\simeq \frac{\lc^2}{D_B}\sim10^{14}\frac{B_{\mu\text{G}}\DrOZZ^2}{T_3}\mbox{ Gyr}\coma
\end{equation}
where $m$, $\mathcal{Z}e$, $v$, $\lambda$, and $r_L= m_i c\, v/(\mathcal{Z}e B)$ are respectively the mass, charge, velocity, mean free path, and Larmor radius of the ion.
We have assumed a magnetic field $B\equiv \BmuG \muG$, and Bohm diffusion, $D_B=c\,k_BT/(16\mathcal{Z}eB)$, where $c$ is the speed of light.
The above inequality assumes magnetized particles scattered inside their Larmor radii.

Such diffusion is too slow to dominate the mixing of ions, and is subdominant to turbulent diffusion and advection.
Turbulence in the ICM can substantially enhance diffusion on the relevant scales, but the resulting timescale depends on the unknown properties of such turbulence.
If one simply assumes that turbulent diffusion alone is responsible for the observed abundance gradients in GCs, then the turbulence timescale would be given by the mixing time \citep[\eg][]{RebuscoEtal2005, GrahamEtAl2006}
\begin{equation}\label{eq:t_turb_Mixing}
\tmix \sim 30\,\DrOZZ^2\mbox{ Gyr}\fin
\end{equation}

However, this estimate stems from cluster cores harboring a spiral structure, associated with bulk flows and ordered magnetic fields.
Thus, the mixing time in Eq.~(\ref{eq:t_turb_Mixing}) may well be dominated by advection in the spiral flow, with turbulent diffusion playing a minor role.
Advection should cause the metallicity below the CF to evolve rapidly, on the advection time $\tadvF$ of the fast flow.
The evolution of the overall metallicity distribution in the cluster should, however, be slower, depending on the volume fraction of the fast flow and on the properties of other flow components. In a simple inflow--outflow model, the spiral structure is dominated by a slow inflow \citep{Keshet2012}, giving an advection time $\tadv\gtrsim 10 \tadvF$.

Spiral CFs provide an efficient isolation between the gas phases above and beneath them, furnished by the strong measured parallel magnetization \citep[][and \NaorEAIP]{ReissKeshet2014}.
Indeed, one infers a suppressed diffusion across these CFs from their sharpness \citep[\eg][]{MarkevitchVikhlinin2007}
suppressed viscosity from the strong shear \citep{KeshetEtAl2010},
and suppressed heat conduction from the sharp temperature jump \citep{EttoriFabian2000}.
The relevant radial lengthscale is therefore the distance between CFs along a given radial ray.
For the typical, logarithmic spiral one finds in observations and simulations, the fractional distance between consecutive CFs is $e^{2\pi\alpha}\simeq 2\mbox{--}4$ \citep[][and \NaorEAIP]{Keshet2012}.
Hence, the relevant lengthscale below a CF is approximately given by the CF radius,
$\lc= f_\Delta\rcf\simeq (1-e^{-2\pi\alpha})\rcf$, where $0.5\lesssim f_\Delta\lesssim 1$.

Figure \ref{fig:timescales_Mixing} shows the timescales discussed above, plotted against radius with respect to the GC center, normalized by the radius $R_{500}$ enclosing a mean density that is $500$ times the critical density of the Universe.
The timescales are computed using the azimuthally averaged, deprojected thermal profiles, derived for each GC in \S\ref{sec:Data Reduction_Mixing} and \S\ref{sec:Spectral Analysis_Mixing}.
To avoid clutter, we plot the radial profile of each timescale only in A2204 (curves), but also include (symbols) each timescale evaluated at the radius of each CF in every GC.
CF radii may be found in Table~\ref{tab:jumps_KZx_Mixing}, and estimates of $R_{500}$ for each GC are provided in Table~\ref{tab:parametersCircs_Mixing}.
In the figure, we adopt $\lc= \Rcf/2$ and $l_{\tiny\text{B},10}=\Rcf/10$; note the strong dependence of some timescales, in particular the conduction time, upon $f_\Delta$.

As the figure shows, for the radial range relevant for our CFs, the dynamical and gravity times are much shorter than the timescales of processes that can modify the entropy and metallicity.
Hence, as anticipated in \S\ref{sec:Introduction_Mixing}, we expect the $K$ jumps and $Z$ drops across these CFs to reveal processes slower than those responsible for maintaining the structure and pressure balance of the spiral.
The fast-flow region just below a discontinuity is interesting, as entropy and metallicity are advected by this flow faster than they can be modified by heat conduction and diffusion.
The $K_i$ and $Z_i$ values can thus provide a handle on the origin of the gas entrained by the fast flow.
Outside the fast flow region, heat conduction and radiative cooling are likely to be faster than advection and particle diffusion, although this depends somewhat on the details of the flow and the turbulence in the spiral.
In these regions, the timescale for evolution in $Z$ is longer than for $K$; in particular, $Z_o$ should provide a better handle on the origin of the gas outside the CF than $K_o$.

Figure \ref{fig:timescales_Mixing} also illustrated that the aforementioned timescales are universal among different GCs.
Namely, each timescale inferred at the radius of each CF in every GC (symbols) agrees with that timescale inferred in A2204 (curves), provided that radii are normalized by $R_{500}$.
This is another testament to the self-similarity of the thermal profiles in GCs, and justifies our joint analyses of different GCs, in particular in {\NaorEAIP} and below.
In particular, in what follows we use the best-fit power law approximation for $\tcool$ obtained using the A2204 data,
\begin{equation} \label{eq:tcool_fit}
\tcool\simeq 6.2(r/0.05R_{500})^{1.6\pm0.1}\Gyr  \fin
\end{equation}

\section{Data reduction and deprojection}\label{sec:Data_Red_Spect_An_Mixing}

Here, we analyze the entropy and metallicity profiles in the GCs A2029, A2142, A2204, and Centaurus, overall and in the vicinity of their CFs.
The deprojection analysis of {\NaorEAIP} is used to infer the $K$ and $Z$ profiles in angular sectors harboring each CF, and to extrapolate each profile to the CF radius, on both sides of the CF.
The
azimuthally-averaged profiles of $K$ and $Z$ in each cluster are independently extracted in this work, from raw \emph{Chanda} data (\S\ref{sec:Data Reduction_Mixing} and \S\ref{sec:Spectral Analysis_Mixing}), and deprojected along the line of sight.
With these results, in (\S\ref{sec:advection_results_Mixing}), we make the assumption that the azimuthally-averaged, deprojected profiles, henceforth referred to as global profiles, approximate the unperturbed ICM.
By comparing a global profile to the values of $K$ and $Z$ on each side of the CF, we are then able to constrain the dynamics.

\subsection{Data reduction}\label{sec:Data Reduction_Mixing}

The GCs A2029, A2142, A2204, and Centaurus were observed with the \emph{Chandra} Advanced CCD Imaging Spectrometer (ACIS).
For details on the observations and our analysis, see \NaorEAIP.
Briefly, the raw data of these GCs are processed by applying the standard event filtering procedure.
Periods of elevated background are identified by using the $2.5 –- 7$ keV light curve in a background region free of cluster emission on the ACIS chip.
We use the standard procedure of modelling the detector readout artefacts and sky background.
Exposure maps are then created using
Alexey Vikhlinin's tools\footnote{http://hea-www.harvard.edu/~alexey/CHAV/}.
These exposure maps account for the position- and energy-dependent variations in effective area and
detector efficiency \citep{WeisskopfEtAl2002}, using a MEKAL model.

The exposure maps and the images of the cleaned event, background, and readout files of the different ObsIDs are co-added in sky coordinates.
After cleaning and co-addition, images of the brightness, $\Sx$, in the $0.8-–7.0$ keV energy band are created by subtracting both background and readout files from the processed event files, and dividing the outcome by the co-added exposure map. For the $\Sx$ maps of the GCs A2029, A2204, A2142, and Centaurus, see {\NaorEAIP}.
In each GC, we exclude point sources from our analysis by visually inspecting the $0.8 -– 7.0$ keV $\Sx$ images.

\subsection{Deprojection process}\label{sec:Spectral Analysis_Mixing}

The deprojected thermal profiles in each angular sector crossing a CF are presented in \NaorEAIP, where the deprojection procedure, assumptions, and underlying 3D models are discussed.
In Table~\ref{tab:jumps_KZx_Mixing} we present, for each of the CFs analyzed in {\NaorEAIP} (except the eastern CF in Centaurus; see discussion in \S\ref{sec:advection_results_Mixing}), the local $K$ and $Z$ values, extrapolated to each side of the CF.
Here, and henceforth, we use sector notations comprised of the cluster identifier with letters designating the cardinal or intercardinal sector direction; for example, A2029SW for the south-western sector in the GC A2029.

As outlined above, in order to gauge the origin of gas found above and below the CF, we require the full-angle azimuthal-averaged, deprojected profiles of entropy and metallicity in the GC.
These so-called global profiles, denoted $K(r)$ and $Z(r)$, are constructed by using the observed data to extract the projected profiles (see \S\ref{subsec:ProjectedProfiles_Mixing}) of brightness $\Sx$, temperature $\TtwoD$, and metallicity $\ZtwoD$, and then fitting them with their projected counterparts (see \S\ref{subsec:DeprojectionProcedure_Mixing}) based on parametric 3D models of $n(r)$, $T(r)$, and $Z(r)$. The global profiles are then identified as the models $K(r)=n(r)^{-2/3}k_B T(r)$ and $Z(r)$, with the best-fit parameters.

\subsubsection{Projected quantities}\label{subsec:ProjectedProfiles_Mixing}

The spectral procedure of extracting the projected profiles of $\Sx$, $\TtwoD$, and $\ZtwoD$ is described in {\NaorEAIP}.
The profiles are extracted in logarithmic radial bins, defined by concentric circular rings.
These bins are centered at the GC center, defined as the X-ray peak. For our analyzed GCs, this position coincides with the position of the cD galaxy to within $\sim10\kpc$ for A2142, and within $\sim1\kpc$ for the other three GCs.
The innermost and outermost radii of each observed property are determined by data availability.

As A2029 and A2142 show a highly elliptical projected morphology, for these two GCs we extract the projected profiles also in elliptical bins.
The elliptical morphology is determined by the position angle, P.A., defined as the orientation of the projected major axis (angle with respect to the west, measured west to north), and the minor-to-major axis ratio, $b/a$.
We determine the parameters P.A. and $b/a$ by fitting isophotes as ellipses, not necessarily centered on the cluster center, and then averaging the best-fit P.A. and $b/a$ values.
For A2029, we thus find averages P.A. $\simeq 111.2^\circ$ and $b/a\simeq 0.75$.
For A2142, we find averages P.A. $\simeq 39.6^\circ$ and $b/a\simeq 0.68$.
As an elliptic substitute for the circular radius, we use the distance from the GC center along the bisector of the (single) CF sector, chosen in {\NaorEAIP} for each cluster.
The projected profiles are finally extracted, using concentric elliptical rings with logarithmic spacing in this elliptical radius.

The azimuthally-averaged profiles of the projected $\Sx$, $\TtwoD$, and $\ZtwoD$ are shown in Figs.~\ref{fig:A2029Sx_Mixing}--\ref{fig:A2029ProjZ_Mixing} for A2029 and in Figs.~\ref{fig:A2142Sx_Mixing}--\ref{fig:A2142ProjZ_Mixing} for A2142, with elliptical bins, and in  Figs.~\ref{fig:A2204Sx_Mixing}--\ref{fig:A2204ProjZ_Mixing} for A2204 and in Figs.~\ref{fig:A3526Sx_Mixing}--\ref{fig:A3526ProjZ_Mixing} for Centaurus, with circular bins.

\subsubsection{Model projection}\label{subsec:DeprojectionProcedure_Mixing}

In order to derive the three-dimensional deprojected thermo-chemical properties $n(\bm{r})$, $T(\bm{r})$, and $Z(\bm{r})$, we require three equations that relate these three quantities to the three observed, projected profiles.
The brightness $\Sx$ is most sensitive to $n$, through the X-ray emissivity $j_x(\vect{r})$ integrated over the line of sight $l$,
\begin{equation}\label{eq:Sx_Mixing}
\Sx=\int j_x\, dl\fin
\end{equation}
The value of the thermal properties $A\in\{T,Z\}$ may be approximated, under certain assumptions, by the weighted average of $A$ with weight $W_A$, taken over the volume $V$ corresponding to a given angular bin $\breve{b}$ and the range of $l$,
\begin{equation}\label{eq:A_EW_Mixing}
A_{proj}(\breve{b})\simeq \frac{\int W_A A \,dV}{\int W_A\,dV}\fin
\end{equation}
Following {\NaorEAIP}, we use the weights $W_T=W_Z=j_x$.

With parametric models for the spatial distributions of $n$, $T$, and $Z$, one can carry out the above three integrals, fit the observed, projected profiles, and thus derive the model parameters.
Next, we describe these three-dimensional spatial models.

\subsubsection{Three-dimensional models}\label{subsec:ModParameters_Mixing}

Deprojection requires a 3D model for each of the thermo-chemical properties, $n$, $T$, and $Z$.
For simplicity, we adopt spherically symmetric models, so each property becomes a function of only the radius $r$.
For clusters A2029 and A2142, which show elliptical projected morphologies, we also consider prolate-spheroidal models, as typically suggested by observations and simulations; furthermore, for simplicity, we assume that the major axis of these prolate distributions lie on the plane of the sky (see discussion in {\NaorEAIP}).
As the parameters P.A. and $b/a$ were already determined based on the projected profiles, each prolate thermo-chemical property too reduces to a function of a single parameter, chosen again as the radius along the CF bisector.

For the electron number density profile, we first attempt a $\beta$-model \citep{CavaliereFuscoFemiano1976,CavaliereFuscoFemiano1978},
\begin{equation}\label{eq:nBetaModel_Mixing}
n(r)=n_0\left(\frac{r_{c}^2+r^2}{r_{c}^2+\rR^2}\right)^{-\frac{3}{2}\beta_n}\coma
\end{equation}
Here, $\beta_n$ is the slope parameter, $r_{c}$ is the core radius, $\rR$ is a reference radius, and
$n_0=n(r_0)$.
If the $\beta$-model yields a negligibly small $r_{c}$ (with respect to the radial data range; as is the case for Centaurus), the model is replaced by a power-law of index $\alpha_n$,
\begin{equation}\label{eq:n_PLaw_Mixing}
n(r) = n_0\left(\frac{r}{\rR}\right)^{\alpha_{n}}\fin
\end{equation}
For A2029, the electron number density indicates a power law, rather than a flat, core.  We thus use an alternative model for this GC: a broken power law \citep[see][]{VikhlininEtAl2006}, where
\begin{equation}\label{eq:n_Vikhlinin1_Mixing}
n(r)=n_0\left(\frac{r}{\rR}\right)^{\alpha_n} \left(\frac{r_{c}^2+r^2}{r_{c}^2+\rR^2}\right)^{-\frac{3}{2}\beta_{n}-\frac{\alpha_n}{2}}\fin
\end{equation}

For the global $T(r)$ and $Z(r)$ profiles, we simply adopt power-law profiles,
\begin{equation}\label{eq:T_PLaw_Mixing}
T(r)=T_0\left(\frac{r}{\rR}\right)^{\alphaT}\coma
\end{equation}
and
\begin{equation}\label{eq:Z_PLaw_Mixing}
Z(r)=Z_0\left(\frac{r}{\rR}\right)^{\alphaZ}\coma
\end{equation}
respectively, with constant normalization factors $T_0$ and $Z_0$, and power-law indices $\alphaT$ and $\alphaZ$.

In A2142 and Centaurus, a power-law model does not provide a good fit to the $T(r)$ data; see for example Fig.~\ref{fig:A2204ProjT_Mixing}.
For these two GCs, we thus use a broken power law model \citep[see][]{VikhlininEtAl2006},
\begin{equation}\label{eq:T_Vikhlinin_Mixing}
T(r)=T_0\left(\frac{r}{\rR}\right)^{\alpha_T} \frac{\left[1+(\rR/\rT)^{\bT}\right]^{\frac{\cT}{\bT}}}{\left[1+(r/\rT)^{\bT}\right]^{\frac{\cT}{\bT}}}\fin
\end{equation}
Here, $\alpha_T$ is the power-law index at small radii, $\cT$ is the correction to the power-law index at large radii, $\rT$ is the transition radius, and $\bT=-2$ \citep[following][]{VikhlininEtAl2006} controls the width of the transition region.
For the prolate-spheroidal ICM distribution in A2142, we relax the $\bT=-2$ constraint, as it gives a non-physical entropy drop at small radii.

In each
GC, we have restricted the radial range of each model according to data availability.
The radial, azimuthally-averaged profiles, of $K$ and $Z$, are presented in Figs.~\ref{fig:A2029K_Mixing}--\ref{fig:A2029Z_Mixing} for A2029, Figs.~\ref{fig:A2142K_Mixing}--\ref{fig:A2142Z_Mixing} for A2142,  Figs.~\ref{fig:A2204K_Mixing}--\ref{fig:A2204Z_Mixing} for A2204, and Figs.~\ref{fig:A3526K_Mixing}--\ref{fig:A3526Z_Mixing} for Centaurus.
The best-fit parameters of the 3D models (Eqs.~\ref{eq:nBetaModel_Mixing}--\ref{eq:T_Vikhlinin_Mixing}) are given in Table~\ref{tab:parametersCircs_Mixing}, for each of the analyzed GCs.

\begin{sidewaystable}
	\caption{Radial GC models} 
	\centering 
	\setlength{\tabcolsep}{0.13em} 
	{\renewcommand{\arraystretch}{1.5}
		\begin{tabular}{ |c |c| c| c| c|  c| c| c| c| c| c| c|c|c|c|c| } 
			\hline 
			GC& ICM&  $n_0[10^{-3}\mbox{cm}^{-3}]$& $T_0[\mbox{keV}]$ &$Z_0[Z_{\sun}]$ & $\rR[\mbox{kpc}]$&$\alpha_n$&  $\beta_n$& $r_c[\mbox{kpc}]$& $\alphaT$&$\bT$&$\cT$&$\rT[\mbox{kpc}]$ &$\alphaZ$&$R_{500}[\mbox{kpc}]$&$M_{200}[10^{14}M_\sun]$\\
			(1) & (2) & (3) & (4) & (5) & (6) & (7)  & (8)& (9)& (10)& (11)& (12)& (13)& (14)& (15)& (16)\\ 
			\hline 
			\multirow{ 2}{*}{A2029}&spherical &$24.7\pm0.3$&$5.8\pm0.3$&$0.7\pm0.1$&\multirow{2}{*}{$31.8\pm0.9$}&$-0.62\pm0.08$&$0.53\pm0.09$&$80\pm30$&$0.22\pm0.05$&\multirow{2}{*}{$-$}&\multirow{2}{*}{$-$}&\multirow{2}{*}{$-$}&$-0.3\pm0.2$&\multirow{2}{*}{$1436^{+79}_{-86}$}&\multirow{2}{*}{$9.1\pm5.6$}\\
			&prolate &$28.4\pm0.3$&$5.7\pm0.2$&$0.66\pm0.09$&&$-0.63\pm0.05$&$0.56\pm0.04$&$100\pm20$&$0.22\pm0.04$&&&&$-0.3\pm0.1$&&\\
			\hline 
			\multirow{2}{*}{A2142}&spherical &$1.58\pm0.01$&$9\pm1$&$0.30\pm0.04$&\multirow{2}{*}{$343.8\pm0.3$}&\multirow{2}{*}{$-$}&$0.66\pm0.01$&$153\pm7$&$4\pm44$&$-2$&$4\pm43$&$1400^{+13000}_{-1400}$&$-0.2\pm0.4$&\multirow{2}{*}{$1500^{+121}_{-100}$}&\multirow{2}{*}{$12.4^{+0.18}_{-0.16}$}\\ 
			&prolate &$1.58\pm0.02$&$9\pm2$&$0.29\pm0.04$&&&$0.68\pm0.03$&$200\pm10$&$0.6\pm0.7$&$-10$&$0.8\pm0.6$&$300\pm200$&$-0.2\pm0.4$&&\\ 
			\hline 
			A2204&spherical&$12.0\pm0.1$&$7.4\pm0.3$&$0.47\pm0.08$&$71.4\pm1.0$&$-1.36\pm0.02$&$-$&$-$&$0.57\pm0.08$&$-$&$-$&$-$&$-0.2\pm0.3$&$1300^{+36}_{-29}$&$7.1^{+3.8}_{-2.6}$\\ 
			\hline 
			Centaurus&spherical&$5.17\pm0.05$&$3.04\pm0.09$&$1.20\pm0.03$&$40.9\pm0.1$&$-1.07\pm0.03$&$-$&$-$&$0.3\pm1.5$&$-2$&$-2\pm2$&$50^{+130}_{-50}$&$-0.68\pm0.06$&$843^{+14}_{-14}$&$1.6^{+0.3}_{-0.2}$\\ 
			\hline 
		\end{tabular}}\label{tab:parametersCircs_Mixing}\vspace{0.2cm}
		\begin{tablenotes}
\item	Parameters of models (\ref{eq:nBetaModel_Mixing}--\ref{eq:T_Vikhlinin_Mixing}). Columns: (1) GC name; (2) ICM distribution, spherical or prolate-spheroidal; (3) electron number density at reference radius $\rR$; (4) temperature at $\rR$; (5) metallicity at $\rR$; (6) reference radius; (7) density power-law index; (8) density $\beta$-model index; (9) core radius in $\beta$ model;  (10) temperature power-law index;  (11) transition shape in a broken power-law model; (12) large radii power-law index correction; (13) transition radius in broken power-law model; (14) metallicity power-law index; (15) the radius enclosing a mean density $500$ times the critical density of the Universe \citep[values taken from][]{ReiprichBohringer2002}; and (16) the total mass inside a radius containing a mean density $200$ times the critical density of the Universe. Weak lensing masses are taken
    from \citet{GonzalezEtAl2018} for A2029 \citep[consistent with the X-ray estimate assuming hydrostatic equilibrium, $M_{200}=(8.0\pm1.5)\times10^{14}M_\sun$;][]{WalkerEtAl2012},
    from \citet{MunariEtAl2014} for A2142 \citep[consistent with the X-ray estimate $M_{200}=(14.5\pm0.3)\times10^{14}M_\sun$;][]{TcherninEtAl2016},
    and from \citet{CorlessEtAl2009} for A2204 \citep[consistent with the X-ray estimate $M_{200}=(6.2\pm0.6)\times10^{14}M_\sun$][]{ReiprichEtAl2009}.
    In Centaurus, we use the X-ray estimate of \citet{WalkerEtAl2013}.
Uncertainties specified in columns (3)--(14) pertain to multi-variant confidence levels.
		\end{tablenotes}
\end{sidewaystable}

\MyApJNew{\begin{figure*}[h]
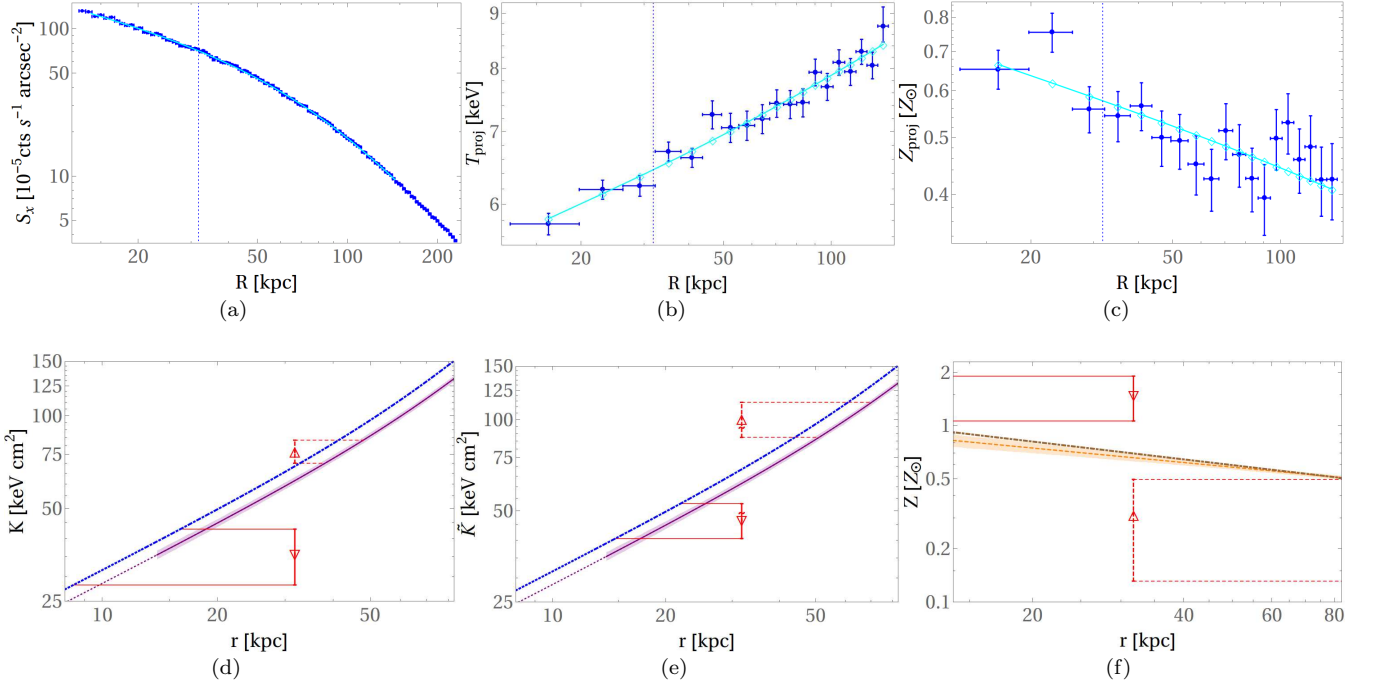

\centering
\subfigure[]
{
\DrawFig{
\includegraphics[width=2.365in]{\myfig{A2029Sx3El.eps}}
}
\label{fig:A2029Sx_Mixing}
}
\hspace{-0.6cm}
\subfigure[]
{
\DrawFig{
\includegraphics[width=2.275in]{\myfig{A2029T2d2El.eps}}}
\label{fig:A2029ProjT_Mixing}
}
\hspace{-0.4cm}
\subfigure[]
{
\DrawFig{\includegraphics[width=2.355in]{\myfig{A2029Z2d2El.eps}}}
\label{fig:A2029ProjZ_Mixing}
}\\
\subfigure[]
{
\DrawFig{\includegraphics[width=2.365in]{\myfig{A2029K4El2.eps}}}
\label{fig:A2029K_Mixing}
}
\hspace{-0.4cm}
\subfigure[]
{
\DrawFig{\includegraphics[width=2.325in]{\myfig{A2029KtdEl2.eps}}}
\label{fig:A2029Ktd_Mixing}
}
\hspace{-0.4cm}
\subfigure[]
{
\DrawFig{\includegraphics[width=2.325in]{\myfig{A2029Z4El2.eps}}}
\label{fig:A2029Z_Mixing}
}
\caption{
Deprojection (top row) and advection (bottom row) analyses of A2029, assuming a prolate-spheroidal ICM distribution.
The profiles of projected, azimuthally-averaged observables (blue error bars) and projected best-fit global (\ie radial, deprojected) models (cyan curves) are plotted against the distance $R$ from the GC center along the bisector of the CF sector.
The top row shows the projected profiles of surface brightness (panel \emph{a}), temperature (panel \emph{b}), and metallicity (panel \emph{c}).
The bottom row shows the global profiles (curves, shaded at $1\sigma$ confidence levels, with extrapolations as dotted curves) of the deprojected entropy (panel \emph{d}; solid purple) and metallicity (panel \emph{f}; dashed orange), as a function of the prolate radius $r$ (chosen equal to $R$; see text).
These profiles are compared (horizontal lines for inferring the $1\sigma$ confidence levels) to the respective values measured below (down-pointed triangles with solid error bars) and above (up-pointed triangle with dashed error bars) the CF, to estimate the radius of origin of the two gas phases.
Also shown are the global profiles deduced by simply assuming a spherical ICM distribution (dot-dashed; blue for $K$ and brown for $Z$).
Panel \emph{e} is identical to \emph{d}, but shows CF values corrected (see Eqs.~\ref{eq:KiTilde_Mixing}--\ref{eq:ZoTilde_Mixing}) against azimuthal variations (tilde-triangles).
}\label{fig:A2029Thermal_Mixing}
\end{figure*}}

\MyApJNew{\begin{figure*}[h]
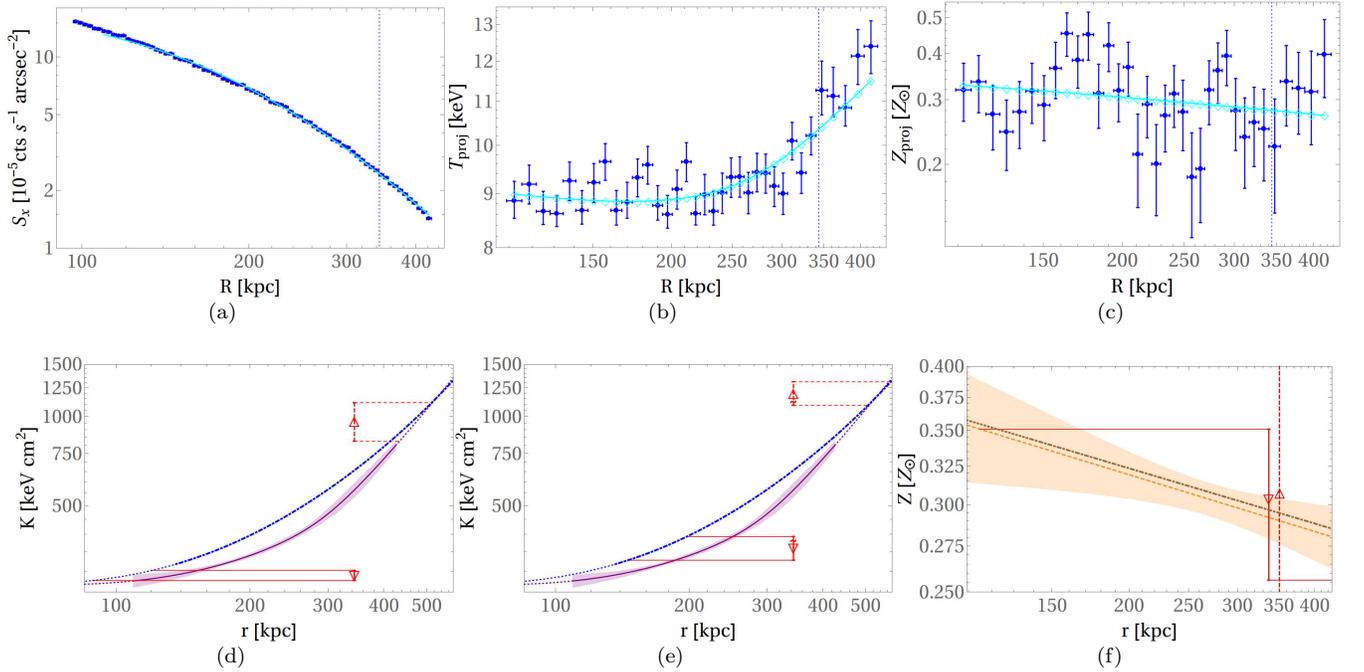

\centering
\subfigure[]
{
\DrawFig{\includegraphics[width=2.295in]{\myfig{A2142Sx2CEl.eps}}}
\label{fig:A2142Sx_Mixing}
}
\hspace{-0.4cm}
\subfigure[]
{
	\DrawFig{\includegraphics[width=2.315in]{\myfig{A2142T2d2BEl.eps}}}
	\label{fig:A2142ProjT_Mixing}
}
\hspace{-0.4cm}
\subfigure[]
{
	\DrawFig{\includegraphics[width=2.375in]{\myfig{A2142Z2d2BEl.eps}}}
	\label{fig:A2142ProjZ_Mixing}
}\\
\subfigure[]
{
	\DrawFig{\includegraphics[width=2.30in]{\myfig{A2142K5El2.eps}}}
	\label{fig:A2142K_Mixing}
}
\hspace{-0.4cm}
\subfigure[]
{
	\DrawFig{\includegraphics[width=2.30in]{\myfig{A2142KtdEl2.eps}}}
	\label{fig:A2142Ktd_Mixing}
}
\hspace{-0.4cm}
\subfigure[]
{
	\DrawFig{\includegraphics[width=2.30in]{\myfig{A2142Z5El2.eps}}}
	\label{fig:A2142Z_Mixing}
}
\caption{Same as \MyApJNew{Figs.}\MyMNRASNew{Figs.}~\ref{fig:A2029Thermal_Mixing}, but for A2142.
In panel \emph{f}, due to the large error bars, symbols for the gas above (below) the CF are slightly offset for visibility to the right (left); the CF radius lies between the two symbols (henceforth).
}\label{fig:A2142Thermal_Mixing}
\end{figure*}}

\MyApJNew{\begin{figure*}[h]
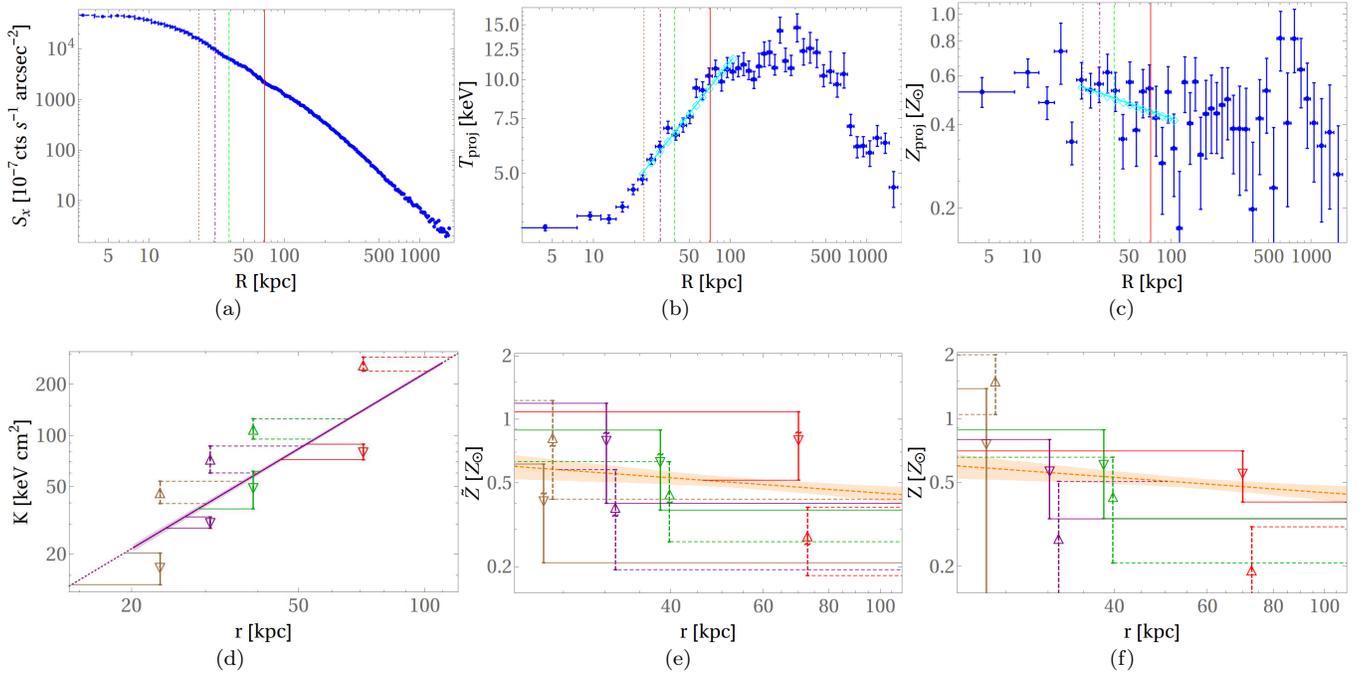

\centering
\subfigure[]
{
\DrawFig{\includegraphics[width=2.35in]{\myfig{A2204Sx3.eps}}}
\label{fig:A2204Sx_Mixing}
}
\hspace{-0.4cm}
\subfigure[]
{
\DrawFig{\includegraphics[width=2.35in]{\myfig{A2204T2d2.eps}}}
\label{fig:A2204ProjT_Mixing}
}
\hspace{-0.4cm}
\subfigure[]
{
\DrawFig{\includegraphics[width=2.34in]{\myfig{A2204Z2d2.eps}}}
\label{fig:A2204ProjZ_Mixing}
}\\
\subfigure[]
{
\DrawFig{\includegraphics[width=2.365in]{\myfig{A2204K4.eps}}}
\label{fig:A2204K_Mixing}
}
\hspace{-0.4cm}
\subfigure[]
{
\DrawFig{\includegraphics[width=2.325in]{\myfig{A2204Ztd.eps}}}
\label{fig:A2204Ztd_Mixing}
}
\hspace{-0.4cm}
\subfigure[]
{
\DrawFig{\includegraphics[width=2.325in]{\myfig{A2204Z4.eps}}}
\label{fig:A2204Z_Mixing}
}
\caption{Same as \MyApJNew{Figs.}\MyMNRASNew{Figs.}~\ref{fig:A2029Thermal_Mixing}, but for A2204, and assuming only a spherical ICM distribution.
This GC harbors four CFs, found (with increasing radii) in sectors A2204S (brown symbols), A2204E (purple), A2204N (green), and A2204W (red).
Here, panel \emph{e} shows metallicity, rather than entropy, values, corrected against azimuthal variations.
}\label{fig:A2204Thermal_Mixing}
\end{figure*}}

\MyApJNew{\begin{figure*}[h]
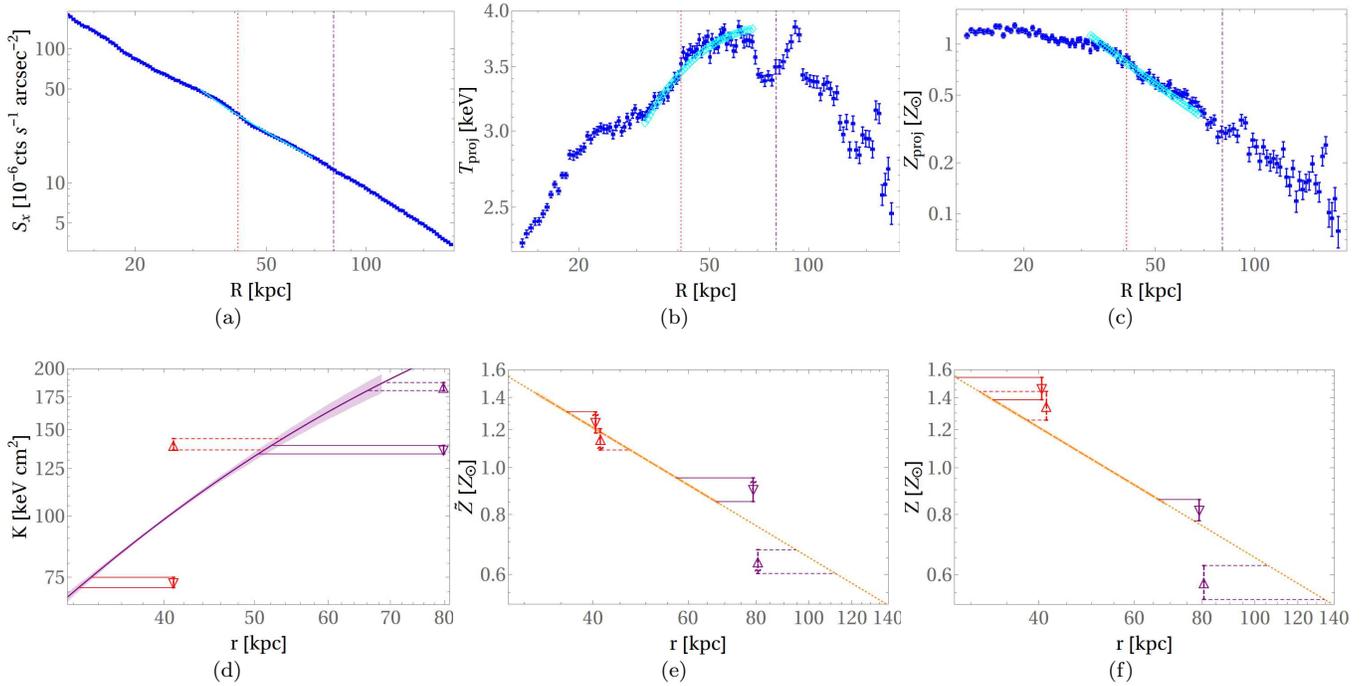

\centering
\subfigure[]
{
\DrawFig{\includegraphics[width=2.35in]{\myfig{A3526Sx2C.eps}}}
\label{fig:A3526Sx_Mixing}
}
\hspace{-0.4cm}
\subfigure[]
{
	\DrawFig{\includegraphics[width=2.34in]{\myfig{A3526T2d2B.eps}}}
	\label{fig:A3526ProjT_Mixing}
}
\hspace{-0.4cm}
\subfigure[]
{
	\DrawFig{\includegraphics[width=2.35in]{\myfig{A3526Z2d2B.eps}}}
	\label{fig:A3526ProjZ_Mixing}
}\\
\subfigure[]
{
	\DrawFig{\includegraphics[width=2.34in]{\myfig{A3526K7.eps}}}
	\label{fig:A3526K_Mixing}
}
\hspace{-0.4cm}
\subfigure[]
{
	\DrawFig{\includegraphics[width=2.355in]{\myfig{A3526Ztd2.eps}}}
	\label{fig:A3526Ztd_Mixing}
}
\hspace{-0.4cm}
\subfigure[]
{
	\DrawFig{\includegraphics[width=2.345in]{\myfig{A3526Z7.eps}}}
	\label{fig:A3526Z_Mixing}
}
\caption{
Same as \MyApJNew{Figs.}\MyMNRASNew{Figs.}~\ref{fig:A2204Thermal_Mixing}, but for Centaurus.
This GC harbors two CFs, found (with increasing radii) in sectors A3526SW (red symbols) and A3526E (purple).
}\label{fig:A3526Thermal_Mixing}
\end{figure*}}

\section{Measuring large-scale advection}\label{sec:advection_results_Mixing}

\subsection{Underlying assumptions}

In this section, we focus exclusively on the deprojected results.
The entropy and metallicity values near the CF differ significantly from their azimuthally-averaged, so-called global counterparts.
Assuming that the spiral structure does not substantially distort the latter, these differences indicate the direction from which the gas, presently found near the CF, has originated, and provide some estimate of the distance it travelled.
Note that the instantaneous flow direction cannot be directly inferred in this method because it may have reversed in the past, as is the case in merger-induced sloshing.

If processes modifying the Lagrangian $K$ and $Z$ values were negligible, we could have used this method to pinpoint the precise radii from which the gas parcels presently above and below the CF have originated.
If, furthermore, the spiral flow was triggered by a sudden, \eg merger, event, then these radii would correspond to the steady-state gas distribution prior to the event.

More realistically, however, we must take into account such processes, shown in \S\ref{sec:TimeScales_Mixing} to modify $K$ over $\gtrsim \Gyr$ timescales, and $Z$ over somewhat longer timescales.
An additional complication is that the flows above and below the CF differ markedly in velocity; the much faster, nearly sonic flow below the CF is, for a fixed travel distance, far younger, and less susceptible to Lagrangian changes in $K$ and $Z$.

One should keep in mind that the mechanism driving the spiral structure has not yet been conclusively determined.
While certain merger events were shown numerically to drive the formation of a spiral, a link between spirals and mergers is challenged by the short spiral lifetime in the presence of cooling, the ubiquity of spiral structures, and by the fast flows inferred below CFs.
Note that, even if the spiral was created by a merger event, it may be sustained over long timescales by other processes, such as AGN bubbles. For a discussion, see \citet{Keshet2012}.

In order to simplify the following discussion, we frame our analysis of the entropy and metallicity distributions with reference to the simplest model of weak, adiabatic sloshing.
In this picture, the Lagrangian $K$ and $Z$ values are approximated as time-independent, the spiral structure is assumed to be violently seeded at some early time, and the initial spatial distributions of $K$ and $Z$ are assumed to be accurately captured by their presently-measured azimuthal averages.

\subsection{Measures of advection}\label{subsec:advect_measures}

Under these assumptions, we may directly compare the entropy values $K_i$ and $K_o$ to the global profile $K(r)$.  Such a comparison yields the radii $\{\riK,\roK\}$ from which the gas $\{$below, above$\}$ the CF has originated, by solving the equations
\begin{equation} \label{eq:RK_Mixing}
K\left(\{\riK,\roK\}\right)=\{K_i,K_o\} \fin
\end{equation}
Analogous origin radii $\{\riZ,\roZ\}$ are inferred by comparing the metallicity values and profile, through
\begin{equation} \label{eq:RZ_Mixing}
Z\left(\{\riZ,\roZ\}\right)=\{Z_i,Z_o\} \fin
\end{equation}
We define radial advection factors $f$ as the fractional distances traveled by the gas according to these origin radii.
Anticipating the flow inside (outside) the CF to originate from smaller (larger) radii, and in order to maintain factors $f>1$ larger than unity, we thus define $\rKircf\equiv \rcf/\riK$ and $\rKorcf\equiv \roK/\rcf$; analogously, $\rZircf\equiv \rcf/\riZ$ and $\rZorcf\equiv \roZ/\rcf$.

In figures \ref{fig:A2029Thermal_Mixing}--\ref{fig:A3526Thermal_Mixing}, panels \emph{d} and \emph{f} show how the origin radii are inferred from the CF values; horizontal lines demonstrate the determination of the $1\sigma$ confidence level on the origin radius of each gas parcel.
Table \ref{tab:jumps_KZx_Mixing} provides the advection factors $f$, evaluated for each CF, based independently on $K$ and on $Z$.

The above measures of advection are sensitive to the values of the global radial profiles, interpolated to the CF radius.
However, $K(\rcf)$ and $Z(\rcf)$ may in practice deviate from the values anticipated at the CF in the absence of advection, especially if
azimuthal variations in the ICM properties are substantial.
We therefore introduce somewhat more robust, mean advection measures, $f_K$ and $f_Z$.
These factors depend on the functional form of the azimuthal averages, but are less sensitive to their precise interpolation --- and in some cases, even extrapolation --- to the CF.

To define $f_K$, assume that the interpolated entropy $K(\rcf)$ gives the anticipated CF value in the absence of advection.
Define effective entropy values below and above the CF, $\tilde{K}_i$ and $\tilde{K}_o$, based on this assumption, while preserving the measured entropy jump $\Kjump=K_o/K_i=\tilde{K}_o/\tilde{K}_i$ across the CF.
Furthermore, let us make the simplifying assumption of a symmetric distribution around $K(\rcf)$, \ie $K(\rcf)/\tilde{K}_i=\tilde{K}_o/K(\rcf)$; partial evidence for such a symmetry is discussed in \S\ref{subsec:advection_K_Mixing}.
We may now define effective origin radii, $\tdriK$ and $\tdroK$, corresponding to $\tilde{K}_i$ and $\tilde{K}_o$, through
\begin{equation}\label{eq:KiTilde_Mixing}
\tilde{K}_i \equiv K(\rcf)/\sqrt{\Kjump} \equiv K(\tdriK) \coma
\end{equation}
and
\begin{equation}\label{eq:KoTilde_Mixing}
\tilde{K}_o \equiv K(\rcf)\sqrt{\Kjump} \equiv K(\tdroK)  \fin
\end{equation}
As asymmetry was assumed between $\tilde{K}_i$ and $\tilde{K}_o$, the mean advection factor of these two phases,
\begin{equation}
f_K\equiv \sqrt{\tdroK/\tdriK}\coma
\end{equation}
is most meaningful.

Analogous quantities are introduced based on the metallicity profiles, namely
\begin{equation}\label{eq:ZiTilde_Mixing}
\tilde{Z}_i \equiv Z(\rcf)\sqrt{\Zjump} \equiv Z(\tdriZ) \coma
\end{equation}
and
\begin{equation}\label{eq:ZoTilde_Mixing}
\tilde{Z}_o \equiv Z(\rcf)/\sqrt{\Zjump} \equiv Z(\tdroZ)  \fin
\end{equation}
The mean advection factor of the two phases, according to metallicity, is similarly defined as
\begin{equation}
f_Z\equiv \sqrt {\tdroZ/\tdriZ} \coma
\end{equation}
in analogy with $f_K$.
The simplifying assumption of symmetry between $\tilde{Z}_i$ and $\tilde{Z}_o$ cannot be tested with present uncertainties.

In figures \ref{fig:A2029Thermal_Mixing}--\ref{fig:A3526Thermal_Mixing}, panel \emph{e} shows such effective origin radii, based on either $\tilde{K}$ or $\tilde{Z}$, depending on the reliability of the $K(r)$ and $Z(r)$ interpolations to the CF.
Horizontal lines demonstrate again the determination of the $1\sigma$ confidence intervals on these effective radii.
Table \ref{tab:jumps_KZy_Mixing} provides the effective advection factors $f$, evaluated for each CF, based on both $\tilde{K}$ and $\tilde{Z}$.

\subsection{Advection analysis: overview}\label{subsec:Advection_analysis_overview_mixing}

In the following, we examine the properties of advection inferred based on $K$ in \S\ref{subsec:advection_K_Mixing}, based on $Z$ in \S\ref{subsec:advection_Z_Mixing}, and based on the combined data in \S\ref{subsec:advection_KZ_Mixing}.
The inferred advection properties are presented in Figures \ref{fig:rkiko_Mixing}--\ref{fig:qrzqrk_Mixing} and in Tables  \ref{tab:jumps_KZx_Mixing}--\ref{tab:jumps_KZy_Mixing}.
In some cases, the global $K$ or $Z$ profile estimated in \S\ref{sec:Data_Red_Spect_An_Mixing} do not cover a sufficiently large radial range, so estimating the gas radius of origin requires extrapolation;
such results are highlighted by a circle in Figures \ref{fig:rkiko_Mixing}--\ref{fig:qrzqrk_Mixing}, and by square brackets in Tables \ref{tab:jumps_KZx_Mixing}--\ref{tab:jumps_KZy_Mixing}.

In A2204, four CFs were analyzed, and the global profiles were inferred for the full CF range, so this cluster provides the most valuable insight into advection in different parts of the cluster.
In Centaurus, the photon statistics are excellent and two CFs were analyzed, but the global profiles, interrupted by substructure, do not reach the radius of the outer, eastern CF (A3526E), which thus requires extrapolation.
Therefore, in this CF we exclude $\Myfi$ and $\Myfo$ from the statistical analysis, but retain the more robust, $\qrK$ and $\qrZ$ factors.
In A2029 and A2142, we were able to analyze only one CF sector in each cluster.
Additional deprojected CFs from the literature are used as well, but only the entropy profiles are available.

\subsection{Advection inferred from entropy}\label{subsec:advection_K_Mixing}

Inspection of Figs.~\ref{fig:A2029Thermal_Mixing}---\ref{fig:A3526Thermal_Mixing} indicates that for all seven CFs that can be compared to their respective global profiles without extrapolation,
\begin{equation}
K_i<K(\rcf)<K_o \coma
\end{equation}
in agreement with the assumption of gas just below (above) the CF originating from a smaller (larger) radius.
The corresponding radial factors $\rKircf$ and $\rKorcf$, plotted in Fig. \ref{fig:rkiko_Mixing} against the normalized CF radius, can thus be used to gauge the flow toward the CF.

\MyApJNew{\begin{figure}[h]}
\MyMNRASNew{\begin{figure}}
	\centering
	\DrawFig{\includegraphics[width=9.0cm,trim={0.3cm 0 0 0}, clip]{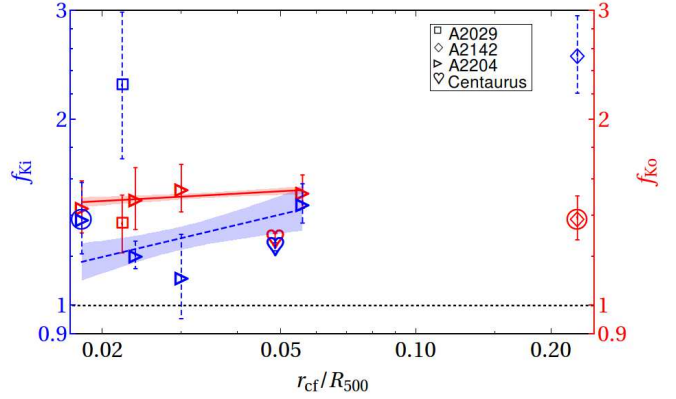}}
	\caption{Advection factors below (left axis; blue symbols with dashed error bars and curves) and above (right axis; red; solid) the seven CFs with non-extrapolated $K(\rcf)$, plotted against the normalized CF radius $\Rcf/R_{500}$.
	Spherical (prolate) ICM models are used for A2204 and Centaurus (A2029 and A2142).
	Best-fit power-law models are provided (curves with shaded $1\sigma$ confidence levels; see text) for the four CFs in A2204.
	Advection factors that require extrapolation are marked by a circle (henceforth).
    The horizontal dotted line corresponds to no advection.
	}
	\label{fig:rkiko_Mixing}
\end{figure}

For the eighth CF, in A3526E, $K(\rcf)$ cannot be determined without extrapolation; here, $\rKircf$ and $\rKorcf$ are distorted and excluded from the figure, whereas $f_K$ may still be meaningful (see Fig. \ref{fig:rfkz_Mixing}).
We include $\rKorcf$ in A2142 although it requires extrapolation for the prolate, albeit not spherical, model, because the extrapolation is mild and to a region where both models are well-behaved and in agreement.
Similarly, we include the mildly-extrapolated $\rKircf$ in A2204. Extrapolated data are circled in the figure.

Unlike the fast flow below the CF, the slow gas above the CF is susceptible to radiative cooling and to the compensating heating mechanism.
Therefore, the radial factor $\rKorcf$ is unlikely to retain memory of the full advection of gas above the CF, or its origin before sloshing ensued.
Instead, if cooling and heating are assumed to be balanced, $\rKorcf$ only gauges the advection over the recent cooling time, and should not vary much among CFs.
Indeed, we find that $1.3\lesssim\rKorcf\lesssim1.5$, with little, $\sigma(\rKorcf)\simeq 0.08$ dispersion
among the seven CFs; the best fit is \begin{equation}
\barrKorcf=1.33\pm0.04 \quad (\chi^2_\nu\simeq2.1) \coma
\end{equation}
where $\chi^2_\nu\equiv \chi^2/\nu$ is the reduced chi-squared value \citep{Pearson1900}, and $\nu$ is the number of degrees of freedom.
Without the low value of $\rKorcf$ in Centaurus with its small uncertainty, this becomes $\barrKorcf=1.47\pm0.03 \quad (\chi^2_\nu\simeq0.4)$.

The small, nearly constant, $\rKorcf\simeq 1.3$ radial factor of the slow phase, consistent with reflecting only the recent advection within the last cooling time, directly identifies this phase as a slow-cooling inflow.
An opposite, outflow scenario seems highly contrived, as it would require radiative cooling to be over-compensated by a heating mechanism, raising the entropy of the gas as it moves outward to a level higher than the surrounding, fine-tuned instead with the environment at a radius $\sim 1.3$ time larger.
A stationary phase is similarly unlikely to retain a constant $\rKorcf$ in the presence of cooling.
It was already pointed out \citep{Keshet2012} that the phase above the CF is likely to be an inflow.
The scaling of the inflow velocity, discussed in \S\ref{sec:Discussion_Mixing}, is shown to be consistent with a simple inflow--outflow model \citep{ReissKeshet2015}.

In contrast to the slow inflow above the CF, the nearly sonic, $\Macht\sim0.8$ gas below the CF cannot radiatively cool over radial factors $f_i\lesssim 10$.
Unless this gas is rapidly mixed or heated, it can thus retain its entropy over a long timescale, so one expects substantial $\rKircf$ variations among CFs.
Indeed,
\begin{equation}
1.1\lesssim\rKircf\lesssim2.5
\end{equation}
shows a significant, $\sigma(\rKircf)\simeq 0.35$, dispersion
among CFs in different clusters.
In general, it is more natural to identify a nearly sonic, persistent flow as an outflow \citep{Keshet2012, ReissKeshet2015}, but the direction of the fast spiral flow below the CF was not conclusively identified until now.
Taking into account the radially rising $K(r)$ profile of this phase, its fast velocity, and the limited range of $\rKircf\lesssim2.5$, we can identify this phase as an outflow undergoing rapid heating or mixing; an opposite, inflow scenario would require rapid non-radiative cooling, which seems unlikely.

It is interesting to examine the behavior of the advection factors within a GC.
The mean values for the four CFs in A2204 are $\barrKorcf=1.51\pm0.02$ ($\chi^2_\nu\simeq0.1$) with little,
$\sigma(\rKorcf)\simeq 0.04$ dispersion,
and a somewhat smaller, $\barrKircf=1.28\pm0.07$ ($\chi^2_\nu\simeq1.9$)
with $\sigma(\rKircf)\simeq 0.1$.
Excluding the extrapolated A2204S CF leaves these results, $\barrKorcf=1.53\pm0.02$ ($\chi^2_\nu\simeq0.04$)
and  $\barrKircf=1.27\pm0.08$ ($\chi^2_\nu\simeq2.7$).
To further test the uniformity in $\rKorcf$ inside this cluster, we fit it by a power law model $\rKorcf\propto \rcf^b$, giving a nearly flat $b=0.04\pm0.03$ ($\chi^2_\nu\simeq0.09$). A TS test \citep[][]{Wilks1938,MattoxEtAl1996} comparing this model with a uniform, $\rKorcf=\const$, null hypothesis, gives $\sim 0.4\sigma$, so the uniform model cannot be rejected.
Similarly testing $\rKircf$ indicates a distribution consistent with uniformity inside the cluster. Fitting a power-law model, $\rKircf\propto \rcf^b$ yields $b=0.2\pm0.1$ ($\chi^2_\nu\simeq1.3$), so a mild increase with radius is possible.
The homogeneous null hypothesis, when compared to this power-law model, gives a TS value of $\simeq1.8\sigma$ confidence, so cannot be rejected. Similarly, a Pearson linear correlation between $\ln(\rKircf)$ and $\ln(\Rcf)$ gives $\corl=0.4^{+0.4}_{-0.6}$, again consistent with uniformity.

The small dispersion of $\rKircf$ inside A2204 and its large dispersion among different GCs suggests that the advection factors may be an inherent property of the GC.
It is not obvious that this should be the case. In merger-induced sloshing, for example, the advection properties are likely to be sensitive to the parameters of the merger, more than to the parameters of the main GC. It is therefore interesting to correlate the advection factor within a GC with other properties of the GC.

In Fig.~\ref{fig:fkiM200_Mixing}, we plot $\rKircf$ against a characteristic virial mass of the GC.  For this purpose, we use $M_{200}$, defined as the total mass inside a radius $R_{200}$ containing a mean density $200$ times the critical density of the Universe.
Mass estimates are chosen, where possible, based on weak lensing (references provided in Table \ref{tab:parametersCircs_Mixing}).
In Centaurus, no weak lensing estimate is available, to our knowledge, so we adopt an estimate of $M_{200}$ derived from X-ray observations assuming hydrostatic equilibrium.
In A2029, A2142, and A2204, the masses inferred from weak lensing and from X-rays are found to be consistent with each other, indicating that using the X-ray-based mass in Centaurus is reasonable.
A Pearson linear correlation test between $\ln(\rKircf)$ and $\ln(M_{200})$ for our sample of seven
CFs (\ie excluding A3526E), yields the correlation coefficient $\corl=0.5^{+0.1}_{-0.2}$, positive at the $\sim2.3\sigma$ confidence level.
Fitting a power-law model $\rKircf\propto M_{200}^{b}$ yields the best-fit $b=0.14\pm0.07$ ($\chi^2_\nu\simeq5.1$).
A TS test for this power-law model, with respect to the null hypothesis of no mass dependence, $\barMyfKi=1.26\pm0.04$ ($\chi^2_\nu\simeq5.9$), rejects the null hypothesis at the $\sim3.1\sigma$ confidence level.
These results suggest that the advection factor $\rKircf$ is indeed an inherent property of the GC, sub-linearly scaling with the virial mass.

\MyApJNew{\begin{figure}[h]}
\MyMNRASNew{\begin{figure}}
	\centering
	\DrawFig{\includegraphics[width=8cm]{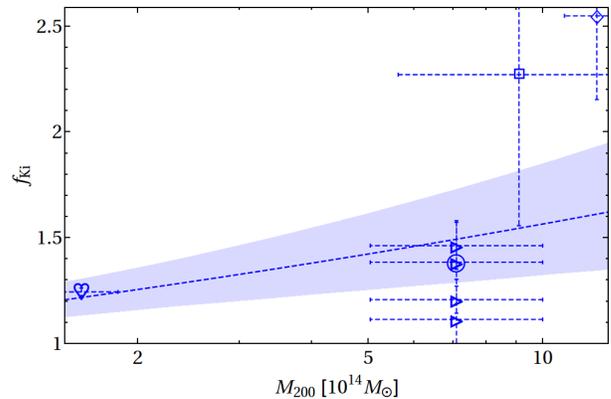}}
\caption{The advection factor $\rKircf$ below CFs, plotted against the virial mass $M_{200}$ of the GC (symbols), along with best-fit power-law (solid curve, with shaded $1\sigma$ confidence level).
}\label{fig:fkiM200_Mixing}
\end{figure}

As $\rKorcf$ is approximately constant, while $\rKircf$ increases with GC mass, it is interesting to compute the ratio $\rKircf/\rKorcf$, and examine the validity of the symmetry assumption used to define $f_K$.
In our best resolved CF, A3526SW, we find comparable, $\rKircf=1.24\pm0.02$ and $\rKorcf=1.28\pm0.03$, so $\rKircf/\rKorcf=0.97\pm0.03$.
In A2204, on average (weighted by statistical errors; henceforth) we find $\langle\rKircf/\rKorcf\rangle=0.89\pm0.06$
($\chi^2_\nu\simeq 1.0$).
The determination of $\rKircf$ in the A2204S CF requires extrapolation; excluding it leaves the result, $\langle\rKircf/\rKorcf\rangle=0.87\pm0.07$ ($\chi^2_\nu\simeq1.3$), nearly unchanged.
In A2029, we find $\rKircf/\rKorcf=1.7\pm0.6$, and in A2142, we find $\rKircf/\rKorcf=1.8\pm0.3$.
It is unknown if $\rKircf/\rKorcf$ remains near unity or becomes small for low-mass clusters.

We conclude that $\rKircf$ and $\rKorcf$ are fairly similar, in particular in our low mass GCs.
It is thus useful to examine the advection factor $\qrK$, computed as defined in Eqs.~(\ref{eq:KiTilde_Mixing}) and (\ref{eq:KoTilde_Mixing}).
This quantity provides a less noisy and more robust measure of the advection, resilient to some extent to azimuthal variations and extrapolation; in particular, we may now consider also the A3526E sector.
The advection factors $\qrK$ are plotted in Fig.~\ref{fig:rfk_Mixing} against $\Rcf/R_{500}$ for the eight analyzed CFs.
As expected based on the $\rKircf$ and $\rKorcf$ behavior, $\qrK$ too shows no significant variations inside a cluster.
For instance, the best-fit power-law model $\qrK\propto \Rcf^{b}$ gives $b=0.09\pm0.06$ for the four CFs in A2204, and $b=-0.09\pm0.07$ for the two CFs in Centaurus, both results being consistent with a constant $\qrK$.

\MyApJNew{\begin{figure*}[h]
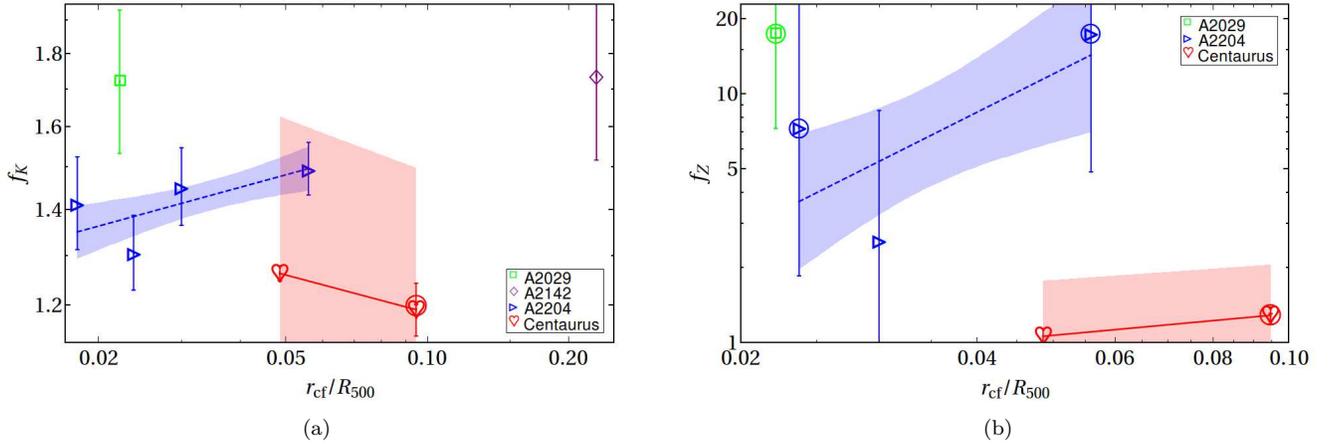

\centering
\subfigure[]
{
\DrawFig{\includegraphics[width=3.45in]{\myfig{rfk2.eps}}}
\label{fig:rfk_Mixing}
}
\subfigure[]
{
	\DrawFig{\includegraphics[width=3.45in]{\myfig{rfz2.eps}}}
	\label{fig:rfz_Mixing}
}
\caption{
Mean advection factors $\qrK$ (left panel) and $\qrZ$ (right panel) of the newly analyzed CFs, plotted where available (symbols as in Fig.~\ref{fig:rkiko_Mixing}) against the normalized CF radius $\Rcf/R_{500}$.
Best-fit power-law models (curves with shaded $1\sigma$ confidence levels) are shown for A2204 (dashed blue), and for the two CFs in Centaurus (solid red).
}\label{fig:rfkz_Mixing}
\end{figure*}}

\MyApJNew{\begin{figure*}[h]
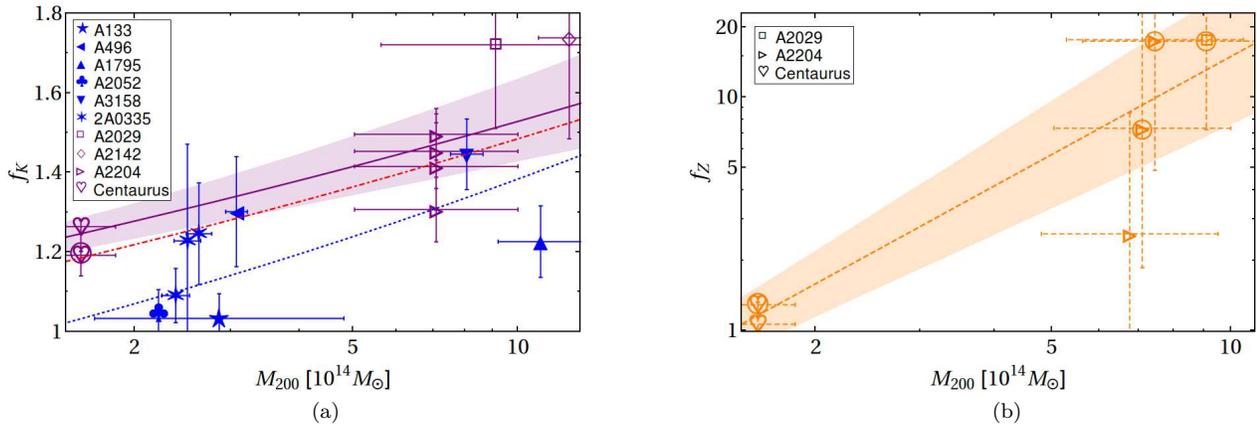

\centering
\subfigure[]
{
\DrawFig{\includegraphics[width=3.45in]{\myfig{rkzM200Comb2fk2.eps}}}
\label{fig:fkM200_Mixing}
}
\subfigure[]
{
	\DrawFig{\includegraphics[width=3.45in]{\myfig{rkzM200Comb2fz2.eps}}}
	\label{fig:fzM200_Mixing}
}
\caption{Same as Fig.~\ref{fig:rfkz_Mixing}, but with $\qrK$ (left panel) and $\qrZ$ (right panel) now plotted against GC mass $M_{200}$, and including $\qrK$ values also for previously deprojected CFs from the literature (filled symbols).
Best-fit power-law models for $\qrK$ are shown for samples of eight newly deprojected CFs (solid purple curve, with shaded $1\sigma$ confidence level), eight previously deprojected CFs (dotted blue), and the joint sample (dot-dashed red).
The best-fit power-law model for $\qrZ$ is seen to be steeper (dashed orange, with shaded $1\sigma$ confidence level).
For presentation purposes, masses of two A2204 CFs are slightly offset.
Data references for previously deprojected CFs: A133 \citep{RandallEtAl2010}, A496 \citep{TanakaEtAl2006}, A1644 \citep{JohnsonEtAl2010}, A1795 \cite{MarkevitchEtAl2001}, A2052 \citep{dePlaaEtAl2010}, A3158 \citep{WangEtAl2010}, 2A0335 \citep{SandersEtAl2009}, and $M_{200}$ \citep{ReiprichBohringer2002,SimionescuEtAl2017}.\label{fig:fM200_Mixing}}
\end{figure*}}

As for $\rKircf$, we also examine a possible correlation of $\qrK$ with $M_{200}$, plotted against each other in Fig.~\ref{fig:fkM200_Mixing}.
A Pearson linear correlation test between $\ln(\qrK)$ and $\ln(M_{200})$ for our sample of eight CFs yields $\corl=0.7^{+0.1}_{-0.2}$, positive at the $\sim3.1\sigma$ confidence level.
Fitting a power-law model
yields the best-fit
\begin{equation}
\qrK\propto M_{200}^{0.11\pm0.03} \quad (\chi^2_\nu\simeq1.0) \fin
\end{equation}
A TS test for this power-law model, with respect to the null hypothesis of no mass dependence, rejects the null hypothesis at the $\sim5.3\sigma$ confidence level.
These results are consistent with the above analysis of $\rKircf$, suggesting that the mean-advection factor $\qrK$ is an inherent property of the GC, sub-linearly scaling with the virial mass.

In order to improve the statistics of our CF analysis, we supplement it with eight CFs from six GCs that were previously deprojected in the literature, and are classified as high-quality CFs in {\NaorEAIP}.
In the absence of a deprojected global profile, we assume that each of these GCs shows the characteristic, linear entropy profile, $K(r)\propto r$, typically inferred in GCs and galaxy groups \citep[][and references therein]{ReissKeshet2015}.
This approximation introduces some error because the entropy power-law has considerable dispersion around unity; for example, we obtain $K(r)\propto r^{0.91\pm0.02}$ in A2029, $K(r)\propto r^{0.66\pm0.04}$ in A2142, $K(r)\propto r^{1.47\pm0.03}$ in A2204, and $K(r)\propto r^{1.36\pm0.01}$ in Centaurus.

For the sample of these eight previously deprojected CFs, on their own accord,
the Pearson correlation test between $\ln(\qrK)$ and $\ln(M_{200})$ gives $\corl=0.5^{+0.2}_{-0.3}$,
positive at the $\sim1.7\sigma$ confidence level.
A power-law model fit $\qrK\propto M_{200}^{b}$ gives $b=0.16\pm0.04$ ($\chi^2_\nu\simeq1.6$), which is TS-test favored over a mass-independent model ($\barqrK=1.17\pm0.05$; $\chi^2_\nu\simeq3.5$) at the $\sim3.9\sigma$ significance level.

For the joint sample of 16 newly and previously deprojected CFs, the Pearson correlation coefficient is $\corl=0.6^{+0.1}_{-0.1}$, positive at the $\sim4.0\sigma$ significance level.
A power-law model is best fit with a slope $b=0.12\pm0.03$ ($\chi^2_\nu\simeq6.8$).
The null hypothesis of mass-independent $\barqrK=1.27\pm0.03$ ($\chi^2_\nu\simeq4.9$) is TS-test rejected with respect to this model at the $\sim5.7\sigma$ confidence level. We conclude that the mass-dependence of $\qrK$ is robust.

\subsection{Advection inferred from metallicity}\label{subsec:advection_Z_Mixing}

Constraining the advection using metallicity is conceptually easier than using entropy, because cooling has no effect, but in practice is more difficult, because the uncertainties in $Z$ are larger and the global $Z(r)$ profile is flatter, typically $Z\sim r^{-0.3}$ \citep{SandersonEtAl2009}.
As a result, the uncertain $Z_i$ and $Z_o$ values are mapped, using $Z(r)$, onto origin radii $\riZ$ and $\roZ$ that are both highly uncertain and typically lie beyond the measured global profile, thus necessitating an extrapolation, as seen in Figs.~\ref{fig:A2029Thermal_Mixing}--\ref{fig:A3526Thermal_Mixing}.

Consequently, with the present data, it is preferable to treat $\riZ$ and $\roZ$ with caution.
Instead, we study the mean advection factor $\qrZ$, corresponding to the effective origin radii $\tdriZ$ and $\tdroZ$.
The underlying assumption of symmetry between $Z_i/Z(\rcf)$ and $Z(\rcf)/Z_o$ cannot be rigorously established, although it appear to qualitatively hold in some cases (see Figs. \ref{fig:A2029Z_Mixing} and \ref{fig:A2204Ztd_Mixing}).

Two CFs, A2204S and A2142NW, are excluded from the metallicity analysis.
The A2204S sector suffers from a combination of a previously unidentified, non-modelled, nearby CF (at $r\simeq55\kpc$), leaving only two radial bins above the deprojected CF, with metallicity oscillations that preclude a reliable fit.
The A2142NW sector shows an exceedingly large uncertainty in $Z_o$, due to the poor statistics at this peripheral radius, with very low density found above the CF.

The remaining sample includes six CFs, all showing $\qrZ>1$, consistent with the sloshing picture.
Four out of these six CFs require extrapolation, as their $\qrZ$ factors are large and the $Z(r)$ profile is fairly flat.
Figures \ref{fig:rfkz_Mixing}--\ref{fig:qrzqrk_Mixing} thus show different properties of $\qrZ$ in the six-CF sample, alongside the $\qrK$ properties of the larger sample.
We are unable to supplement the present analysis with previously deprojected CF metallicity drops from the literature, becasue the only relevant CFs are already deprojected here.

While $\rKorcf$ and possibly $\rKircf$ (and thus $\qrK$) variations inside a GC are limited by nonthermal cooling and heating, $\qrZ$ may vary considerably inside a GC.
Fitting the three remaining CFs in A2204 with a power-law model $\qrZ\propto \Rcf^b$ (see Fig. \ref{fig:rfz_Mixing}) suggests a radially rising profile, $b=1.6\pm1.7$ ($\chi^2_{\nu}\simeq0.7$), but the uncertainty is substantial.
TS-testing this model with respect to a homogeneous $\qrZ$ null hypothesis gives $\sim0.7\sigma$, so the null hypothesis cannot be rejected.
Similarly, a Pearson linear correlation test between $\ln(\qrZ)$ and $\ln(\Rcf)$ yields a positive but highly uncertain correlation coefficient $\corl=0.6^{+0.3}_{-1.2}$.
Qualitatively similar results are obtained in Centaurus, where a similar power-law fit indicates $b=0.3\pm0.2$, slightly positive but still consistent with a uniform advection factor.
We conclude there is preliminary evidence for radially-increasing $\qrZ(r)$, but the poor statistics prevent a meaningful measurement at this stage.

A Pearson linear correlation test between $\ln(\qrZ)$ and $\ln(M_{200})$ in the six-CF sample (Fig. \ref{fig:fzM200_Mixing}) yields the correlation coefficient $\corl=0.7^{+0.2}_{-0.3}$, positive at the $\sim2.0\sigma$ confidence level.
Best fitting a power-law model
yields a significantly positive best-fit slope,
\begin{equation}
\qrZ\propto M_{200}^{1.4\pm0.4} \quad (\chi^2_\nu\simeq0.4) \fin
\end{equation}
A TS test for this model with the null hypothesis of a mass-independent $\qrZ$ rejects the null hypothesis at the $\sim4.2\sigma$ confidence level.
These results indicate that $\qrZ$ is a rather strong function of virial mass, rising significantly faster than its entropy-based counterpart, $\qrK(M_{200})$.

This correlation manifests in large metallicity-based advection factors in massive clusters, greatly exceeding their entropy-based counterparts.
Among the CFs in A2204, we find $\barqrZ=7\pm3$ ($\chi^2_{\nu}\simeq0.6$) on average, noticeably larger than their entropy-based counterparts, $\barqrK=1.43\pm0.04$ ($\chi^2_{\nu}\simeq1.1$)
on average.
An even larger, $\qrZ=17^{+24}_{-10}$ is found in A2029, far exceeding the entropy-based estimate, $\qrK=1.7\pm0.2$.
For the two CFs in the low-mass Centaurus, on the other hand, we find a modest, $\barqrZ=1.2\pm0.1$ ($\chi^2_{\nu}\simeq3.3$), comparable to the entropy-based, $\barqrK=1.26\pm0.02$ ($\chi^2_{\nu}\simeq1.7$).

In massive clusters, the tension between the large $\qrZ$ factors and the modest $\qrK$ factors supports the above conclusion, reached in \S\ref{subsec:advection_K_Mixing} by analyzing $\rKircf$ and $\rKorcf$.
In these clusters, substantial entropy differences associated with large-scale advection appear to be largely dissipated by non-thermal processes, in particular radiative cooling for the slow inflow, and heating or mixing for the fast outflow.
These processes may play a lesser role in lower-mass clusters such as Centaurus, where $\qrZ$ and $\qrK$ are found to be in approximate agreement.
In such a case, the approximately equal $\rKircf$ and $\rKorcf$ in Centaurus may reveal an unbiased, symmetric advection.

\subsection{Entropy-based vs. metallicity-based advection}\label{subsec:advection_KZ_Mixing}

The preceding discussion suggests a strong correlation between the advection factors inferred from entropy and those inferred from metallicity.
Such a correlation is indeed expected in any model in which advection plays an important role.
In particular, recall that in the adiabatic, merger-driven sloshing model, one expects $\qrK\simeq \qrZ$.

In Fig.~\ref{fig:qrzqrk_Mixing}, $\qrK$ is plotted against $\qrZ$ for the sample of six CFs for which both quantities are available.
A Pearson linear correlation test between $\ln(\qrZ)$ and $\ln(\qrK)$ indicates a positive correlation coefficient $\corl=0.7^{+0.2}_{-0.3}$, marginally significant at the $\sim1.9\sigma$ confidence level.
A power-law model
is best-fit by a shallow slope,
\begin{equation}
\qrK\propto\qrZ^{0.08\pm0.03} \quad (\chi^2_\nu\simeq1.1) \coma
\end{equation}
associated with the large variation in $\qrZ$ vs. the small variation in $\qrK$.
A TS test of this model against a null hypothesis of constant $\qrK$, rejects the null hypothesis at the $\sim4.9\sigma$ confidence level.

\MyApJNew{\begin{figure}[h]}
\MyMNRASNew{\begin{figure}}
	\centering
	\DrawFig{\includegraphics[width=9.6cm]{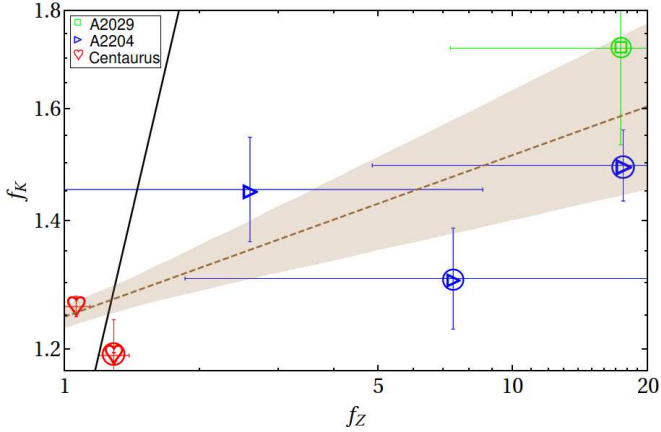}}
	\caption{Mean advection factors $\qrK$ and $\qrZ$, plotted against each other.
	Symbols are the same as in Fig.~\ref{fig:rfkz_Mixing}.
	The best-fit power-law model (dashed brown curve with shaded $1\sigma$ confidence level) is much shallower than the $\qrK=\qrZ$ of adiabatic sloshing (black solid curve).
    }
	\label{fig:qrzqrk_Mixing}
\end{figure}

These results show a positive correlation between the advection factors inferred from entropy and from metallicity, with the latter becoming very large in massive clusters.
The positive correlation supports a model in which CFs are a manifestation of large-scale advection, due to some dynamical process reflected both in entropy and in metallicity.
However, the mass dependence of $\qrZ$ is much stronger than that of $\qrK$, leading to a shallow $\qrK(\qrZ)$ that is inconsistent with the $\qrK\simeq \qrZ$ behavior (solid line in Fig.~\ref{fig:qrzqrk_Mixing}) anticipated in the adiabatic, merger-driven sloshing model.

\begin{sidewaystable}
\caption{CF entropy and metallicity, and the implied advection
} 
\centering 
{\renewcommand{\arraystretch}{1.7}%
\begin{tabular}{|c|c|c|c|cc|cc|cc|cc|} 
\hline 
GC& ICM&sector & $\rcf$[kpc]&$K_i[\keV \cm^{2}]$&$K_o[\keV \cm^{2}]$&$\log_{10}\rKircf$&$\log_{10}\rKorcf$&$Z_i[Z_{\sun}]$&$Z_o[Z_{\sun}]$&$\log_{10}\rZircf$&$\log_{10}\rZorcf$\\
(1) & (2) & (3) & (4) & (5)& (6)& (7)& (8)& (9)& (10)& (11)& (12)\\ 
\hline 
\multirow{2}{*}{A2029}&spherical& \multirow{2}{*}{SW} &\multirow{2}{*}{$31.8\pm0.9$}&\multirow{2}{*}{$36\pm7$}&\multirow{2}{*}{$77\pm7$}&$\left[0.42^{+0.15}_{-0.12}\right]$&$0.07^{+0.05}_{-0.05}$&\multirow{2}{*}{$1.49\pm0.42$}&\multirow{2}{*}{$0.31\pm0.18$}&$>0.36$&$>0.42$\\
&prolate& &&&&$0.36^{+0.15}_{-0.12}$&$0.13^{+0.05}_{-0.05}$&&&$>0.40$&$>0.66$\\
\hline 
\multirow{2}{*}{A2142}&spherical& \multirow{2}{*}{NW}&\multirow{2}{*}{$343.8\pm0.3$} &\multirow{2}{*}{$290\pm11$}&\multirow{2}{*}{$968\pm144$}&$\left[0.51^{+0.08}_{-0.05}\right]$&$0.13^{+0.04}_{-0.05}$&\multirow{2}{*}{$0.30\pm0.05$}&\multirow{2}{*}{$0.31\pm0.28$}&$0.07^{+>}_{-<}$&$-0.10^{+>}_{-<}$\\
&prolate& & &&&$0.41^{+0.08}_{-0.05}$&$\left[0.13^{+0.04}_{-0.05}\right]$&&&$0.07^{+>}_{-<}$&$-0.10^{+>}_{-<}$\\
\hline 
\multirow{4}{*}{A2204}&\multirow{4}{*}{spherical}
& S&$23.4\pm0.8$&$17\pm4$&$47\pm7$&$\left[0.14^{+0.07}_{-0.06}\right]$&$0.16^{+0.04}_{-0.05}$&$0.76\pm0.62$&$1.52\pm0.48$&$>0.06$&$<-0.06$\\
& &E&$30.8\pm0.6$&$31\pm2$&$73\pm13$&$0.08^{+0.02}_{-0.02}$&$0.17^{+0.05}_{-0.06}$&$0.57\pm0.15$&$0.27\pm0.23$&$0.05^{+>}_{-<}$&$>0.55$\\
& &N&$39.0\pm1.0$&$49\pm12$&$110\pm15$&$0.05^{+0.07}_{-0.08}$&$0.19^{+0.04}_{-0.04}$&$0.61\pm0.27$&$0.43\pm0.22$&$>0.29$&$>0.45$\\
& &W &$71.4\pm1.0$&$80\pm8$&$263\pm25$&$0.17^{+0.03}_{-0.03}$&$0.18^{+[0.03]}_{-0.03}$&$0.55\pm0.15$&$0.19\pm0.11$&$0.36^{+>}_{-<}$&$>0.19$\\
\hline 
\multirow{2}{*}{Centaurus}&\multirow{2}{*}{spherical}
& SW&$40.9\pm0.1$ &$73\pm2$&$140\pm4$&$0.095^{+0.007}_{-0.007}$&$0.11^{+0.01}_{-0.01}$&$1.46\pm0.08$&$1.35\pm0.09$&$>0.11$&$-0.08^{+0.05}_{-<}$\\
&& E&$79.8\pm0.4$ &$137\pm3$&$184\pm4$&$\left[0.192^{+0.009}_{-0.007}\right]$&$\left[-0.073^{+0.008}_{-0.008}\right]$&$0.82\pm0.04$&$0.58\pm0.05$&$<0.07$&$>0.07$\\
\hline 
\end{tabular}}\label{tab:jumps_KZx_Mixing}\vspace{0.2cm}
\begin{tablenotes}
\item	Columns: (1) GC name; (2) ICM distribution, spherical or prolate-spheroidal; (3) analyzed sector; (4) CF radius; (5) sectorial entropy just below the CF; (6) sectorial entropy just above the CF; (7) inner, entropy-based advection factor, $\rKircf\equiv\rcf/\riK$; (8) outer, entropy-based advection factor, $\rKorcf\equiv\roK/\rcf$; (9) sectorial metallicity just below the CF; (10) sectorial metallicity just above the CF; (11) inner, metallicity-based advection factor, $\rZircf\equiv\rcf/\riZ$; and (12) outer, metallicity-based advection factor, $\rZorcf\equiv\roZ/\rcf$.
An inequality symbol (in a value or in confidence level) indicates that only a lower limit ($>$) or an upper limit ($<$) could be found.
Square brackets designate advection factors that required extrapolation.
\end{tablenotes}
\end{sidewaystable}

\begin{sidewaystable}
\caption{CF entropy and metallicity contrasts, and the implied advection} 
\centering 
{\renewcommand{\arraystretch}{1.5}%
\begin{tabular}{|c|c|c|c|cc|c|c|c|c|} 
\hline 
GC& ICM& sector & $\rcf$[kpc]&$\tilde{K}_i[\keV \cm^{2}]$&$\tilde{K}_o[\keV \cm^{2}]$&$\Kjump$&$\Zjump$&$\log_{10}\qrK$&$\log_{10}\qrZ$\\
(1) & (2) & (3) & (4) & (5)& (6)& (7)&(8)&(9)&(10)\\ 
\hline 
\multirow{2}{*}{A2029}& spherical& \multirow{2}{*}{SW} &\multirow{2}{*}{$31.8\pm0.9$}&$46\pm6$&$100\pm13$&\multirow{2}{*}{$2.2_{-0.4}^{+0.6}$}&\multirow{2}{*}{$5^{+3}_{-2}$}&$0.23\pm0.05$&$\left[1.01\pm0.33\right]$\\
& polate& &&$42\pm6$&$90\pm12$&&&$0.24\pm0.05$&$\left[1.24\pm0.38\right]$\\
\hline 
\multirow{2}{*}{A2142}& spherical& \multirow{2}{*}{NW}&\multirow{2}{*}{$343.8\pm0.3$} &$360\pm33$&$1200\pm110$&\multirow{2}{*}{$3.3_{-0.5}^{+0.6}$}&\multirow{2}{*}{$1.0^{+1.0}_{-0.5}$}&$0.25\pm0.03$&$\left[-0.02\pm0.59\right]$\\
& prolate& & &$309\pm31$&$1030\pm105$&&&$0.24\pm0.06$&$\left[-0.02\pm0.58\right]$\\
\hline 
\multirow{4}{*}{A2204}
&\multirow{4}{*}{spherical}
& S&$23.4\pm0.8$&$16\pm3$&$45\pm7$&$2.8_{-0.7}^{+0.9}$&$0.5^{+0.5}_{-0.2}$&$0.15\pm0.03$&$\left[-0.83\pm0.59\right]$\\
&& E&$30.8\pm0.6$&$27\pm4$&$60\pm8$&$2.2_{-0.4}^{+0.5}$&$2^{+2}_{-1}$&$0.12\pm0.03$&$\left[0.87\pm0.60\right]$\\
&& N&$39.0\pm1.0$&$33\pm4$&$100\pm13$&$3.0_{-0.6}^{+0.8}$&$1.4^{+1.1}_{-0.6}$&$0.16\pm0.03$&$0.41\pm0.52$\\
&& W &$71.4\pm1.0$&$78\pm7$&$255\pm23$&$3.3_{-0.5}^{+0.6}$&$3^{+2}_{-1}$&$0.17\pm0.02$&$\left[1.25\pm0.56\right]$\\
\hline 
\multirow{2}{*}{Centaurus}&\multirow{2}{*}{spherical}& SW&$40.9\pm0.1$ &$74\pm2$&$141\pm3$&$1.92_{-0.08}^{+0.08}$&$1.1^{+0.1}_{-0.1}$&$0.102\pm0.005$&$0.03\pm0.03$\\
&&E&$79.8\pm0.4$ &$\left[185\pm9\right]$&$\left[249\pm13\right]$&$1.35_{-0.04}^{+0.04}$&$1.25^{+0.09}_{-0.08}$&$\left[0.08\pm0.02\right]$&$\left[0.11\pm0.04\right]$\\
\hline 
\end{tabular}}\label{tab:jumps_KZy_Mixing}\vspace{0.2cm}
\begin{tablenotes}
\item	Columns: (1) GC name; (2) ICM distribution, spherical or prolate-spheroidal; (3) analyzed sector; (4) CF radius; (5) entropy just below the CF, corrected for azimuthal variations, $\tilde{K}_i\equiv K(\rcf)/\Kjump^{1/2}$; (6) entropy just above the CF, corrected for azimuthal variations, $\tilde{K}_o\equiv K(\rcf)\Kjump^{1/2}$; (7) CF entropy contrast $\Kjump\equiv K_o/K_i$; (8) CF metallicity contrast $\Zjump\equiv Z_i/Z_o$; (9) entropy-based, mean advection factor $\qrK\equiv(\tdroK/\tdriK)^{1/2}$; and (10) metallicity-based, mean advection factor $\qrZ\equiv(\tdroZ/\tdriZ)^{1/2}$.
Square brackets designate extrapolation.
\end{tablenotes}
\end{sidewaystable}

\section{Summary and Discussion}\label{sec:Discussion_Mixing}

We study the entropy $K$ and metallicity $Z$ distributions around spiral CFs in GCs, in order to measure the advection of the gas and constrain the underlying flow.
We focus on $K$ and $Z$ because they are thought to evolve on timescales (see Fig.~\ref{fig:timescales_Mixing}) much longer than the dynamical time, providing clues that supplement the dynamical constraints inferred previously on pressure balance and shear (\NaorEAIP, and references therein).
The timescales of ICM processes relevant for Lagrangian evolution in $K$ and $Z$, reviewed in \S\ref{sec:TimeScales_Mixing}, show considerable self-similarity among GCs, collapsing to universal profiles if radii are normalized by $R_{500}$.
Such self-similarity warrants a unified study of multiple CFs in several GCs.

We focus on eight CFs that were recently analyzed in {\NaorEAIP}: one CF in A2029 (see Fig.~\ref{fig:A2029Thermal_Mixing}), one in A2142 (Fig.~\ref{fig:A2142Thermal_Mixing}), four in A2204 (Fig.~\ref{fig:A2204Thermal_Mixing}), and two in Centaurus (Fig.~\ref{fig:A3526Thermal_Mixing}).
The deprojected profiles of these CFs all show entropy jumps, and all but one show metallicity drops (bottom panels of Figs.~\ref{fig:A2029Thermal_Mixing}--\ref{fig:A3526Thermal_Mixing}).
Table~\ref{tab:jumps_KZx_Mixing} provides the statistics of $K$ and $Z$ on each side of each CF.

It would be interesting to compare the different values of $K$ and $Z$, on each side of a CF, to their expected values in the absence of a CF.
As a proxy of these putative values, we use the azimuthally-averaged $K$ and $Z$ in the host GC, at the radius of the CF: spherical radius in A2204 and Centaurus, and elliptical radius in the morphologically elliptical A2029 and A2142.

For this purpose, raw \emph{Chandra} data of the four GCs (see \S\ref{sec:Data Reduction_Mixing}) are reduced and analyzed
(see \S\ref{sec:Spectral Analysis_Mixing}) in order to derive the deprojected,
full-angle azimuthally averaged, thermo-chemical profiles in each GC. The averaged, so-called global, profiles of density, temperature, entropy, and metallicity, are shown in Figs.~\ref{fig:A2029Thermal_Mixing}--\ref{fig:A3526Thermal_Mixing}.
Table~\ref{tab:parametersCircs_Mixing} provides the best-fit parameters of the deprojected models (Eqs.~\ref{eq:nBetaModel_Mixing}--\ref{eq:T_Vikhlinin_Mixing}) for these profiles.

The low $K$ and high $Z$ values found below a CF suggest that this gas phase rose from a deeper part of the GC at some time in the past.
Analogously, the high $K$ and low $Z$ values found above the CF suggest that this phase had an earlier peripheral origin.
Such an hypothesis can be tested, for example by correlating the $K$ and $Z$ offsets with each other.
The interpretation of the results is complicated by the very different velocities of the two components, and by the different timescales of Lagrangian $K$ and $Z$ evolution.

To simplify the analysis, we parameterize it in terms of an oversimplified, adiabatic weak sloshing picture.
Here, a steady-state GC is assumed to be perturbed in the past by some violent, \eg merger, event, leading to an ideal flow with no Lagrangian $K$ or $Z$ evolution, that in addition maintains approximately the same global $K$ and $Z$ profiles of the unperturbed gas. In such a model, the $K$ and $Z$ values near the CF would both accurately correspond to the same, so called origin, radius of the gas prior to the perturbation.

Comparing the local $K$ and $Z$ values near the CF to the respective global profiles, we thus derive the
origin radii, $\riK$ and $\riZ$, for the gas below the CF, and origin radii $\roK$ and $\roZ$ for the gas above the CF (see definitions \ref{eq:RK_Mixing}--\ref{eq:RZ_Mixing}).
In the weak adiabatic sloshing model, one expects $\riK=\riZ$ and $\roK=\roZ$; furthermore, simulations \citep[\eg][]{AscasibarMarkevitch2006} suggest that approximately $\rcf/r_i\simeq r_o/\rcf\simeq 1\text{--}2$.
The measured radii and their associated uncertainties are illustrated in panels \emph{d} and \emph{f} of Figs.~\ref{fig:A2029Thermal_Mixing}--\ref{fig:A3526Thermal_Mixing}.

The entropy data, which can be measure to better accuracy than metallicity, indicate radial advection factors $\rKircf\equiv\rcf/\riK$ and $\rKorcf\equiv\roK/\rcf$ (see Fig.~\ref{fig:rkiko_Mixing} and Table~\ref{tab:jumps_KZx_Mixing}) that are approximately uniform inside a cluster.
In the best-studied case of A2204, we find averages $\barrKircf=1.27\pm0.08$ and $\barrKorcf=1.53\pm0.02$, which are approximately equal, with a ratio $\langle\rKircf/\rKorcf\rangle=0.87\pm0.07$ ($\chi^2_\nu\simeq1.3$).
These results indicate large-scale gas advection over factors $>1.5$ in radius, consistent with global sloshing in the cluster.

Moreover, we find that $\rKorcf$ is consistent with being constant not only inside A2204, but also among CFs in different GCs.
Averaging all seven CFs that do not require extrapolation (\ie excluding A3526E) yields $\barrKorcf=1.33\pm0.04$ ($\chi^2_\nu\simeq2.1$).
This result suggests that the entropy of the slow gas phase above the CF is thermally regulated, in particular by radiative cooling, so $\rKorcf$ does not directly reflect the spatial advection factor.
Instead, $\rKorcf$ imposes a lower limit on the advection factor, and gauges the advection of gas over the recent cooling time.
Taking into account the monotonically rising $K(r)$ profile of gas above the CF, we identify this phase as a slow, cooling, spiral inflow.

Let us assume that over timescales longer than $\tcool$, radiative cooling of the slow inflow can be balanced by some compensating heating or mixing mechanism, so its entropy would equilibrate with its surrounding. Under such circumstances, the advection of gas above the CF would pinpoint its origin at a time $\sim \theat\sim \tcool$ ago.
The inflow Mach number may then be estimated as
\begin{equation} \label{eq:MachO}
\Macht_o = \frac{v_o}{c_s} \sim \frac{(\MyfKo-1)6r}{c_s \tcool}\sim 0.01\, r_{100}^{-0.6\pm0.1} T_3^{-1/2} \coma
\end{equation}
where we used the best-fit $t_{cool}(r)$ of Eq.~(\ref{eq:tcool_fit}).
For the four CFs in A2204, in particular, we obtain on average $\bar{\Macht}_o\simeq0.026\pm0.007$.

In contrast to the uniform behavior of $\rKorcf$, we find a substantial dispersion among the $1.1<\rKircf<2.5$ values in different GCs.
The nearly constant $\rKircf$ value of the four CFs inside A2204 suggests that this factor should be viewed as a property of the GC, possibly correlating with the GC parameters. Indeed, we find that $\rKircf$ correlates (at the $\sim3.1\sigma$ TS-test level) with the GC mass $M_{200}$ (see Fig.~\ref{fig:fkiM200_Mixing}), scaling as $\rKircf\propto M_{200}^{0.14\pm0.07}$.

In some GCs, azimuthal variations and data limitations require extrapolation, so we define
in Eqs.~(\ref{eq:KiTilde_Mixing}--\ref{eq:ZoTilde_Mixing}) effective, symmetric, more robust origin radii
$\tilde{r}_K$ and $\tilde{r}_Z$,
illustrated in panels \emph{e} of Figs.~\ref{fig:A2029Thermal_Mixing}--\ref{fig:A3526Thermal_Mixing}.
The geometrical means $\qrK\equiv(\tdroK/\tdriK)^{1/2}$ and $\qrZ\equiv(\tdroZ/\tdriZ)^{1/2}$ of these effective advection factors are provided in Table~\ref{tab:jumps_KZy_Mixing}, and depicted in Figs.~\ref{fig:rfkz_Mixing}--\ref{fig:qrzqrk_Mixing}.
These mean factors provide more robust estimates of the advection, although they cannot distinguish between different advection factors below and above the CF.

Both $\qrK$ (see \S\ref{subsec:advection_K_Mixing}) and $\qrZ$ (\S\ref{subsec:advection_Z_Mixing}) are consistent with being uniform inside a GC, but the best fit for $\qrZ$ prefers, within the poor statistics, a super- (sub)-linearly rising $\qrZ(r)$ profile inside A2204 (Centaurus); see Fig.~\ref{fig:rfkz_Mixing}.
We find a strong correlation between $\qrK$ and $\qrZ$, with coefficient $\corl=0.7^{+0.2}_{-0.3}$, supporting large-scale advection as the common origin of both phenomena.
The scaling $\qrK\propto \qrZ^{0.08\pm0.03}$ ($\chi^2_\nu\simeq1.1$) is inconsistent with adiabatic model (see Fig.~\ref{fig:qrzqrk_Mixing}), indicating substantial cooling of the hot phase and heating or mixing of the cold phase.

The large-scale advection we find depends strongly on the GC mass, as revealed by strong correlations of both $\qrK$ and $\qrZ$ with $M_{200}$ (see Fig.~\ref{fig:fM200_Mixing}).
The entropy jumps in our CF sample indicate a weak but significant mass dependence, $\qrK\propto M_{200}^{0.11\pm0.03}$ ($\chi^2_\nu\simeq1.0$).
This result is supported by a sample of eight high-quality CFs that were previously analyzed in the literature, which alone gives $\qrK\propto M_{200}^{0.16\pm0.04}$ ($\chi^2_\nu\simeq1.6$).
The metallicity-based data show a stronger mass dependence, $\qrZ\propto M_{200}^{1.4\pm0.4}$ ($\chi^2_\nu\simeq0.4$), with large $\qrZ$ in massive clusters.

Unlike entropy, metallicity is insensitive to cooling and heating processes, although its gradients too can be dissipated by mixing.
The metallicity-based factors $\qrZ$ therefore reflect more faithfully the spatial advection.
The advection factors we infer vary between $\barqrZ=1.2\pm0.1$ ($\chi^2_{\nu}\simeq3.3$) in the lower-mass Centaurus, to
$\barqrZ=10^{+4}_{-3}$ ($\chi^2_{\nu}\simeq0.6$) in the massive clusters A2029 and A2204.

The large $\qrZ$ and small $\qrK$ factors found in massive clusters support the aforementioned conclusion, that the entropy-based advection factors are regulated, at least in these clusters, by non-thermal processes.
In particular, in A2204, comparing the large $\qrZ=7\pm3$
with the small $\rKorcf\simeq 1.5$ is consistent with the entropy of the slow inflow being strongly regulated by radiative cooling.

The cold gas below the CF was previously identified as a fast, $\Macht_i\sim0.8$ flow \citep[][\NaorEAIP]{KeshetEtAl2010}.
The direction of this flow, in particular its identification as either an outflow or an inflow, was until now unknown \citep[but see indications for it being an outflow in][which also discuss the three-dimensional structure of the spiral flow]{Keshet2012, ReissKeshet2015}.
Such a fast flow is not substantially affected by radiative cooling within the $\rKircf\lesssim2.5$, entropy-based radial factors.
However, $\rKircf$ is found to be nearly independent of $r$, whereas the cold phase has a nearly linearly rising $K(r)$ profile, thus indicating that some non-thermal mechanism is rapidly modifying the entropy of this cold phase.
Such a mechanism is most likely to heat the gas, by some non-adiabatic mechanism such as gas mixing, immediately identifying the cold phase as an outflow.

The mechanism maintaining a nearly constant, $\MyfKi\simeq 1.3$ factor, must be able to efficiently heat the fast outflow over a short, $\sim(\MyfKi-1)6r$ travel distance.
This corresponds to a short, $\sim 0.3r_{100}T_3^{-1/2}\Gyr$ timescale, suggesting a combination of mixing and heat conduction over the narrow width of the fast flow layer (see Naor et al. 2020, in preparation).

The large $\qrZ$ factors, on their own accord, similarly suggest that the cold gas below the CF is an outflow, as the opposite scenario seems implausible. Indeed, it is difficult to envisage a mechanism collecting very high-metallicity gas from the cluster periphery and channeling it inward toward the center, maintaining a nearly-azimuthal velocity close to the speed of sound.

Moreover, if gas mixing is not substantial, metallicity can pinpoint the origin of the outflow, and a linear $\qrZ\propto r$ should emerge.
The large $\qrZ$ factors found in A2204 and A2029 (albeit not in Centaurus) would thus suggest an outflow stemming from the cD galaxy, especially if these results and the strong $\qrZ(z)$ dependence in A2204 could be securely established with future observations.

The emerging picture of a mixed spiral flow, consisting of a fast outflow below the CF and a slow cooling inflow above the CF, differs from the flow pattern seen in sloshing simulations.
Here, a simulated violent event such as a merger perturbs the core, which subsequently returns to equilibrium, so the low entropy, high metallicity phase below the CF is an inflow.
Such simulations tend to show modest advection factors \citep[$\sim 2$; see \eg][]{AscasibarMarkevitch2006}, that depend more on the details of the merger event and on the time that elapsed since the merger (Gal et al., in prep.), and less on the mass of the host GC.
Additional challenging observations include the ubiquity of spirals, suggesting stability against radiative cooling and little dependence on the timing of a recent merger, and outflow velocities which are nearly sonic and approximately azimuthal.

Mixed spiral inflows--outflows were modelled previously in \citet{Keshet2012, ReissKeshet2015}, and were shown to reproduce some properties of GCs, in particular stability against radiative cooling and the observed, linearly rising entropy profile.
Interestingly, the $v_o\propto r^{-0.6\pm0.1}$
scaling inferred in Eq.~(\ref{eq:MachO}) for the inflow velocity is marginally consistent with the $r^{-3/7}$ scaling anticipated in such a simple inflow--outflow model \citep{ReissKeshet2015}.
However, a central simplifying assumption of these models is the approximation of the fast outflow as adiabatic; the rapid heating of this component, presumably triggered by mixing, was only crudely incorporated.

Our results are challenged by the small size of the CF sample and by deprojection uncertainties.
For instance, the two most massive GCs in our sample are also those showing an elliptical morphology and thus modeled as prolate spheroids, so any geometric dependence of the advection factors would bias our results.
Our results can be tested and better modeled by expanding the CF sample, utilizing future deprojection methods more suitable for the non-spherical gas distribution, non-adiabatic numerical simulations, and deep X-ray observations.
In particular, space telescopes such as \emph{XRISM} and \emph{ATHENA}, expected to become operational in the next decade, should provide far better constraints on the spiral structures and their underlying flow.

\acknowledgements
We are greatly indebted to M. Markevitch for his extensive help, patience, and hospitality.
We thank Y. Gal for helpful discussions.
This research was supported by the Israel Science Foundation (grant No. 1769/15), by the IAEC-UPBC joint research foundation (grant 300/18), and by the Ministry of Science, Technology \& Space, Israel, and has received funding from the GIF (grant I-1362-303.7/2016).

\bibliography{\mybib{Clusters}}
	
\label{lastpage}	
\end{document}